\documentclass[amssymb, amsmath, nofootinbib,tightenlines,nobibnotes, 10pt, prd,aps,showpacs]{revtex4}
\usepackage{graphicx}
\usepackage{dcolumn}
\usepackage{bm}

\usepackage{color}
\usepackage{times}

\newcommand{\dx}{{d}x}
\renewcommand{\d}{{d}}
\newcommand{\e}{{e}}

\newcommand{\pa}{\partial}
\newcommand{\ie}{{i.e.\ }}
\newcommand{\opl}{{\rm\bf \hat{{\bf L}}}}
\newcommand{\opL}{{\rm\bf \hat{L}}}
\newcommand{\Opl}{{\rm\bf \hat{L}}}
\newcommand{\lab}[1]{}
\newcommand{\Or}{{\cal O}}
\newcommand{\be}{\begin{equation}}
\newcommand{\ee}{\end{equation}}

\def\ellop{{\rm\bf \hat{L}}}
\def\lag{{\mathcal{L}}}
\def\sn{\mathop{\text{sn}}}
\def\cn{\mathop{\text{cn}}}
\def\dn{\mathop{\text{dn}}}
\def\zb{\ensuremath{\bar{z}}}
\def\xii#1{\xi^{(#1)}}
\def\xij#1#2{{\xi^{(#1)}_{#2}}}
\renewcommand\Re{\mathop{\text{Re}}}
\def\PutFigure#1{
    \noindent\hfil\includegraphics[scale=.4,angle=-90]{#1}
}

\begin{document}

\title{Negative radiation pressure exerted on kinks}

\author{P\'eter Forg\'acs$^{1,2}$, \'Arp\'ad Luk\'acs$^{1}$ and Tomasz Roma\'nczukiewicz$^{3}$}
\affiliation{ {$^{1}$MTA RMKI Budapest 114, PO Box 49, H-1525, Hungary}\\
{$^{2}$Laboratoire de Math\'{e}matiques et Physique Th\'{e}orique CNRS\\
 Universit\'{e} de Tours, Parc de Grandmont, 37200 Tours, France}\\
{ $^{3}$Institute of Physics, Jagiellonian University,\\
Reymonta 4, 30-059 Cracow, Poland}
}

\begin{abstract}
The interaction of a kink and a monochromatic plane wave
in one dimensional scalar field theories is studied.
It is shown that in a large class of models the radiation pressure
exerted on the kink is negative,
i.e.\ the kink is {\sl pulled} towards the source of the radiation.
This effect has been observed by numerical simulations in the $\phi^4$ model,
and it is explained by a perturbative calculation
assuming that the amplitude of the incoming wave is small.
Quite importantly the effect is shown to
be robust against small perturbations of the $\phi^4$ model.
In the sine-Gordon (sG) model the time averaged
radiation pressure acting on the kink turns out to be zero.
The results of the perturbative computations in the sG model are shown to be in
full agreement with an analytical solution corresponding to the superposition of a sG kink
with a cnoidal wave.
It is also demonstrated that the acceleration of the kink satisfies Newton's law.
\end{abstract}
\pacs{11.10.Lm, 11.27.+d}
\maketitle

\section{Introduction}

It is by now universally accepted that spatially localized
solutions of nonlinear equations (solitary waves, particle-like objects) are of great importance
in several areas of physics, we refer
to some of the recent monographs \cite{Scott}, \cite{Manton-Sutcliffe}, \cite{Dauxois}.
The study of localized solutions in various systems in a single spatial dimension
(i.e.\ in $1+1$ dimensional space-time) has proved to be quite fruitful in hydrodynamics,
condensed matter physics and also for particle physics as a theoretical laboratory.
A special class of non-linear equations admits ``genuine'' soliton solutions
that retain their shape even after interactions, this feature
distinguishes them from more generic particle-like objects. One can mention the sine-Gordon (sG)
equation as a prototype example admitting ``genuine'' soliton solutions.
It is of considerable interest both from a theoretical point of view and
for physical applications to study the interactions of these spatially localized objects
(not only of ``genuine'' solitons).
There is a well developed framework
for soliton perturbations in integrable and in near integrable systems
\cite{Fogel-etal}, \cite{Kivshar-Malomed},
where many powerful methods of integrable systems are available.
In generic models, however, one has to resort to
various perturbation techniques and in general also to numerical simulations.
The motion of a localized object when subjected to a force
has been shown to be governed by Newton's law for non-relativistic velocities
\cite{Fogel-etal}. It has turned out that in some cases, such as the sG model, deformation
effects are also important and can lead to deviations from Newton's law
\cite{Reinish}, \cite{Rice}.

In this paper we study the motion of particle-like solutions in $1+1$
dimensional scalar field theories (these objects are commonly referred to as ``kinks'')
under the influence of an incident wave.
In one of the most studied scalar models, in the ``$\phi^4$'' theory,
we have found that the kink
starts to accelerate in {\sl the direction} of the incoming wave.
This effect has been first observed in Ref.\ \cite{trom0}, in the present work
we study it in much more detail both numerically and analytically.
The unusual behaviour of the kink can be interpreted as being
caused by a ``negative radiation pressure'' exerted on it.

We have computed the force exerted on the kink by the radiation in a generic model
in perturbation theory, assuming that the amplitude, $A$, of the incoming radiation is small.
The leading order force exerted on a kink by a wave coming from the right
(i.e.\ from $x=\infty$) is then given as $F= -A^2q^2\left|{\mathbf R}\right|^2$,
where $q$ is the wave number of the incident wave,
and ${\mathbf R}$ is the reflection coefficient.
This means that the kink is pushed back by the radiation as expected.
Now in a class of theories containing among others the $\phi^4$ and the sG models,
the leading order reflection coefficient is zero, ${\mathbf R}=0$.
In such cases
the force is determined by higher order terms,
in fact in the next order the force is $F\sim{\cal O}(A^4)$.
The basic physical effect responsible for the negative radiation pressure can then
be understood as follows.
For small enough amplitudes of the incoming radiation,
the kink of the $\phi^4$ theory is {\sl transparent} to the waves to first order in $A$.
This is due to the reflectionless nature of the effective potential in the kink background.
Because of the nonlinearities, during the interaction higher frequency waves are also generated.
Some of the energy of the incoming wave is transferred to higher frequency
(mostly double frequency) waves. These higher frequency waves carry more momentum
than the incoming one.
In the case of the $\phi^4$-type models the amplitude of the transmitted waves
with double frequency is larger than that of the reflected ones.
This way a momentum surplus is created behind the kink, thus pushing it forward.
For large enough amplitudes the radiation pressure becomes positive.

Comparing the perturbative result with numerical simulations, reasonably good agreement
has been found for both the non-relativistic acceleration of the kink,
and for the force acting on it.
Also for small enough amplitudes of the incident wave the kink starts to accelerate
according to Newton's law.

There is no such negative radiation pressure exerted on the kink in the sG model,
although the sG kink is also transparent to the waves.
As it turns out the sG kink is transparent to all orders in $A$ (this special feature of the sG
model is due to its integrability). In fact there is an analytical
solution found in Ref.\ \cite{Shin} corresponding to the nonlinear superposition of a kink
and a traveling wave, which corresponds precisely to the problem we study.
We have compared our perturbative results for the sG kink with
this analytical solution and the perfect agreement found serves as a good test
of our approach.

In our context the negative radiation pressure appears because the kink is
transparent to small amplitude incoming waves to
first order in perturbation theory. Since this happens only in rather special
cases, the structural stability of the effect should be addressed.
We have demonstrated the robustness of the negative radiation
pressure on the example of a generic perturbation of the $\phi^4$
model.
For an arbitrarily small perturbation of
the $\phi^4$ theory, to first order in the amplitude of the incoming wave
the radiation pressure exerted on the kink becomes positive, just as expected in a generic theory.
The higher order contributions tend to compensate this however, and
we have found that even for non-infinitesimal
perturbations of the $\phi^4$ model there exists a critical amplitude of the incoming wave,
above which the radiation pressure becomes negative again.
This robustness of the effect makes it worthwhile for further
studies.

The organization of the paper is as follows. In Section II we introduce the models
and present the first order calculation of the radiation pressure exerted
by an incoming wave on the kink.
In Section III we discuss the higher order
perturbative calculations of the force and of the kink's acceleration.
In Section IV these results are applied to the $\phi^4$ and to the sG model,
where an analytic formula for the force exerted on the kink is derived,
and we also elucidate the physical reasons for the negative radiation pressure.
In Section V the basic setup for the numerical simulations used is given
and the analytical results are compared to those obtained by the numerical simulations.
In Section VI the structural stability of the negative radiation pressure with respect to perturbations
of the Lagrangian is studied.

Most of the computational details are relegated to three Appendices.
In Appendix A some details of the higher order perturbative computations are given,
In Appendix B we perform a suitable expansion
of the analytical solution of the sG kink on a cnoidal wave pertinent to our problem.

\section{Formulation of the problem}

\subsection{The models considered}
We consider the following class of scalar theories in 1+1 space-time dimensions specified
by the Lagrangian
\footnote{our conventions are: $g_{\mu\nu}=\textrm{diag}(1,-1)$, $\pa_0f=\pa_tf=\dot{f}$,
$\pa_1f=\pa_xf={f}'$}:
\begin{equation}
  \label{eq:lagrangian}
    \lag = \frac{1}{2}\partial_\mu\phi\partial^\mu\phi - U(\phi)\,,
\end{equation}
where the self-interaction potential, $U(\phi)$, is assumed to possess at least
two degenerate minima (vacua), denoted by $U(\phi_{vac})$).
The equation of motion obtained from the Lagrangian \eqref{eq:lagrangian} can be written as
\begin{equation}\label{eq:eq1}
\ddot{\phi}-\phi''+U'(\phi)=0\,,
\end{equation}
The class of theories \eqref{eq:lagrangian} admits static, finite energy solutions of
the field equations \eqref{eq:eq1}. These solutions
interpolate between two vacua of $U(\phi)$ and are commonly referred to as
a kink or antikink. A kink (or antikink), $\phi_s(x)$, is given explicitly by the
following formula
\begin{equation}\label{eq:kinkdef}
x-x_0=\pm\int_{\phi(x_0)}^{\phi(x)}\frac{d\phi}{\sqrt{2U(\phi)}}\,.
\end{equation}
In this paper we shall mostly work with the ``$\phi^4$''
where $U(\phi)=(\phi^2-1)^2/2$, resp.\ with the ``sine-Gordon'' (sG)
where $U(\phi)=1-\cos\phi$ models. Note that in the case of the
$\phi^4$ model,
$\phi_{vac}=\phi(\pm\infty)=\pm1$, and $\phi_{vac}=\phi(\pm
\infty)=2\pi, 0$ for the sG equation.

The corresponding kink solutions are well known:
\be
\phi_s=\tanh x\,,\ {\rm in\ }\phi^4\,,\ {\rm resp.}\ \phi_s=4\arctan \exp(-x)\,,\ {\rm in\ the\ sG\ model}\,.
\ee
These kinks are well (exponentially) localized, and their position is conveniently defined
by the zero of $\phi_s(x)$, which also corresponds to the maximum of their energy density.

The energy-momentum tensor of the scalar field theory \eqref{eq:lagrangian} is given as:
\begin{equation}
T_{\mu\nu}=\partial_\mu\phi\partial_\nu\phi-g_{\mu\nu}{\cal L}\,,
\end{equation}
whose components are spelled out explicitly for later convenience:
\begin{eqnarray}
T_{00}&=&{\cal E}=\frac{1}{2}\dot{\phi}^2+\frac{1}{2}\phi'^2+U(\phi),\\
T_{01}&=&T_{10}=-{\cal P}=\phi'\dot{\phi},\label{eq:noether}\\
T_{11}&=&\frac{1}{2}\dot{\phi}^2+\frac{1}{2}\phi'^2-U(\phi),
\end{eqnarray}
where ${\cal E}$ and ${\cal P}$ are the energy and momentum densities, respectively.
The energy-momentum conservation laws $\partial_\mu T^\mu_{\;\;\nu}=0$ can be written as:
\begin{eqnarray}
\partial_t{\cal E} &=& \partial_x\left(\phi'\dot{\phi}\right)\,,\label{eq:ZasE}\\
\partial_t{\cal P} &=& -\frac{1}{2}\partial_x\left(\dot{\phi}^2+\phi'^2-2U(\phi)\right)\,. \label{eq:ZasP}
\end{eqnarray}

\subsection{Interaction of the kink with radiation}
The physical problem we wish to study is the interaction of a kink with an
incoming (scalar) radiation. We assume that at least in the case when the
radiation can be considered as a small perturbation, it is a reasonable approximation
to treat the kink as a particle accelerating under the force coming from the
radiation pressure exerted by the radiation.
More precisely we consider the problem that a monochromatic
wave, $\xi(t,x)$, coming from the right (from $x=\infty$) is incident on an initially static
kink at $x=0$.
The incident wave $\xi(t,x)$ itself reduces asymptotically to a plane wave, i.e.
\be\label{xi}
\xi(t,x\to\infty)\to A\Re\{ e^{i(\omega t+qx)}\}\,,
\ee
where $A$ is the asymptotic amplitude.
The dispersion relation between the frequency, $\omega$, and the wave number, $q$,
is easily read off from Eq.\ \eqref{eq:eq1}:
\begin{equation}
-\omega^2+q^2+U''(\phi_{vac})=0\,.
\end{equation}
In order to allow for
a perturbative solution of the time dependent problem, we shall assume that the amplitude of
the incoming wave, $A$, is sufficiently small and
expand the solution of the nonlinear wave equation \eqref{eq:eq1}
in power series of $A$:
\begin{equation}
\phi=\phi_s+\xi=\phi_s+A\xi^{(1)}+A^2\xi^{(2)}+\cdots\,,
\label{eq:post1}
\end{equation}
To first order in $A$, the ``radiation'', $\xi^{(1)}$ satisfies a linear
wave equation in the background of the kink:
\begin{equation}\label{eq4}
{\ddot{\xi}^{(1)}}-{\xi^{(1)}}''+U''(\phi_s(x))\xi^{(1)}\equiv\ddot{\xi}^{(1)}+\opl\xi^{(1)}=0\,,
\end{equation}
where $U''(\phi_s(x))$ corresponds to the potential
of the Schr\"odinger-type operator $\opl=-d^2/dx^2+U''(\phi_s(x))$.
This potential
can be written explicitly for the ``$\phi^4$'' and for the sG models as
\be\label{potentials}
U''(\phi_s^{\phi^4})=4-\frac{6}{\cosh^2x}\ \hbox{in the}\ \phi^4\quad {\rm resp.}\quad
U''(\phi_s^{\rm sG})=1-\frac{2}{\cosh^2x} \hbox{ in the sG model}\,.
\ee
Introducing
\begin{equation}\label{xi1}
\xi^{(1)}=\frac{1}{2}\left(e^{i\omega t}\eta_q(x)+e^{-i\omega t}\eta_{-q}(x)\right):=
\Re\{ \eta_{q}(x)e^{i(\omega t)}\}\,,
\end{equation}
equation (\ref{eq4}) can be separated,
where the $\eta_q$ are eigenfunctions of the Schr\"odinger operator $\opl$,
\begin{equation}\label{eta}
\opl\eta_q:=\left(-\frac{\d^2}{\dx^2}+U''(\phi_s(x))\right)\eta_q(x)=\omega^2\eta_q(x)\,.
\end{equation}
As it is well known from elementary quantum mechanics, for potentials tending to zero
for $|x|\to\infty$,
the asymptotic forms of the scattering eigenfunctions are:
\begin{eqnarray}
\eta_q(x\rightarrow+\infty)&=&e^{iqx}+{\mathbf R}e^{-iqx}\,,\label{line1}\\
\eta_q(x\rightarrow-\infty)&=&{\mathbf T}e^{iqx}\,,\label{line2}
\end{eqnarray}
where ${\mathbf R}$ and ${\mathbf T}$ are the reflection and transition coefficients.
The first order solution, $\xi^{(1)}$, corresponds to the incoming radiation field, which
reduces asymptotically to a monochromatic plane wave coming from the right to the kink.
Let us remind the reader at this point, that
the kink in any translationally invariant theory possess
a discrete eigenfunction with eigenvalue $\omega=0$ (a zero mode),
which is called the translational mode.
In some cases the Schr\"odinger operator $\opl$ also possess other discrete eigenstates
for $\omega>0$.
There is one such a discrete (or internal) mode in the ``$\phi^4$'' model for $\omega_d=\sqrt{3}$,
and there is none in sG model.

From the energy conservation law, Eq. \eqref{eq:ZasE} it is easily seen that to first order
the change of the total energy of the
kink+radiation system in a box of size $2L$, is
\be\label{energy-cons}
\partial_t E = \int\limits_{-L}^L dx \partial_t{\cal E}=
A^2{\xi^{(1)}}'{\dot{\xi^{(1)}}}\mid^L_{-L}\,.
\ee
Averaging in time over a period, $T=2\pi/\omega$,
one finds
$$
\langle\partial_t E|^L_{-L}\rangle_T=A^2q\omega(1-|{\mathbf T}|^2-|{\mathbf R}|^2)/2\,,
$$
where $\langle F\rangle_T$ denotes the average of the quantity $F$ in time (over a period $T$).

Assuming that the size of the box is sufficiently large and that the
radiation of the kink itself can be neglected, the energy contained
in the box is conserved, and then it follows that
\begin{equation}\label{cons1}
|{\mathbf R}|^2+|{\mathbf T}|^2=1\,.
\end{equation}
The rate of change of the total momentum in the box $[-L,L]$
can be identified with the total force exerted on
the the system of kink+radiation in its inside.
To linear order in perturbation theory from Eq.\ \eqref{eq:ZasP} this force is found to be
\begin{equation}\label{F1}
\left\langle\left.\partial_t P\right|_{-L}^L\right\rangle_T=F^{(2)}=\frac{1}{2}A^2q^2\left(-1-\left|{\mathbf R}\right|^2
+\left|{\mathbf T}\right|^2\right)=-A^2q^2\left|{\mathbf R}\right|^2\,.
\end{equation}

We now show that for not too long times the kink obeys Newton's law under the action
of the force given by \eqref{F1}.
In order to define the acceleration of the kink we make the (usual) assumption
that for small enough velocities one can neglect deformation and other effects
and approximate the moving kink by
$\phi(x,t)=\phi_s(x-X(t))$. This approximation corresponds to introducing simply
a collective coordinate for the position of the kink.
Therefore for small enough displacements, $\phi(x,t)\approx\phi_s(x)-X(t)\phi_s'(x)$.
In fact $\eta_t:=\phi_s'(x)$ is nothing but
the translational zero mode of the kink, which is orthogonal to all other
linearized (internal and radiation) modes of the operator $\opl$.
The leading order displacement of the kink is then easily obtained:
\begin{equation}
  \label{eq:genPOSproj}
  X(t)=-\frac{( \eta_t| \xi)}{( \eta_t | \eta_t)}\,,
\end{equation}
where  $( f |g)$ denotes the natural Hilbert space scalar product.
The acceleration, $a$ to leading order can then be calculated as follows
\begin{equation}
  \label{eq:genACCproj}
  a^{(n)}=\ddot{X}(t)=-\frac{( \eta_t | \ddot{\xi}^{(n)} )}{( \eta_t | \eta_t)}  =
-\frac{( \eta_t |{\ddot\xi}^{(n)}+ \ellop \xi^{(n)} )}{( \eta_t | \eta_t)}\,,
\end{equation}
where ${{\xi}}^{(n)}$ is the
lowest order approximation i.e.\ the smallest $n$
for which the acceleration is non zero. To compute higher order corrections to the acceleration
is nontrivial, since in our approximation the time dependence of the
kink has been encoded in the single collective coordinate, $X(t)$,
whereas one has to take into account distortion, radiation, etc effects.
In the generic case discussed above
$n=2$. Therefore to lowest order in PT the force \eqref{F1} is quadratic in the amplitude,
$A$, and clearly according to our definition \eqref{eq:genACCproj}
the acceleration of the kink is also ${\cal O}(A^2)$.
Then the second order perturbative solution, ${{\xi}}^{(2)}$, is needed.
A not too difficult computation (see Eqs.\ \eqref{eq:accel1}, \eqref{eq:boundary1}
in Appendix A) yields:
\begin{equation}
m_s a^{(2)}=-A^2(\eta_t|\ddot{\xi}^{(2)}_0) =
\frac{A^2}{4}( U'''(\phi_s)\eta_{q}\eta_{-q}|\eta_t)=-A^2q^2\left|{\mathbf R}\right|^2\,,
\label{eq:accel0}
\end{equation}
where the relation $(\eta_t|\eta_t)=m_s$ has been used.
The method to project unto the translational mode to compute the kink's acceleration
was also used in Ref.\ \cite{KisShnir} where the dynamics
of a kink in the $\phi^4$ model with a perturbed potential was studied.
One sees that to leading order in PT the acceleration of the kink, $a^{(2)}$,
is indeed given by Newton's law, $F^{(2)}=m_sa^{(2)}$.
In particular one can see that the kink is pushed back under the action of the force
coming from the radiation pressure as expected.
Therefore it is consistent to identify the time averaged momentum flow in the box
Eq.\ \eqref{F1}, with the {\sl total force} acting {\sl on the kink} to leading order.
When the radiation field can be treated as a small
perturbation the effect of the momentum flow on the radiation field itself can be
neglected as a first approximation.
One would then expect that for small enough amplitudes of the incident radiation,
it is a reasonably good approximation to the solution of equation \eqref{eq:eq1},
that an initially static kink starts to accelerate as a non-relativistic
particle of mass $m_s$.
It is natural to expect that other effects, such as the radiation by the kink,
its distortion, etc.\  show up only in higher orders.
We remind the reader that identifying the force exerted by the radiation on the kink
with the total momentum flow in the box is valid
only to leading order (and after time averaging), and also for relatively short time intervals.
The force (\ref{F1}) is quadratic both in the amplitude of the incoming wave and in
the reflection coefficient, in complete analogy to
the well known radiation pressure in classical electrodynamics.

\begin{figure}
\PutFigure{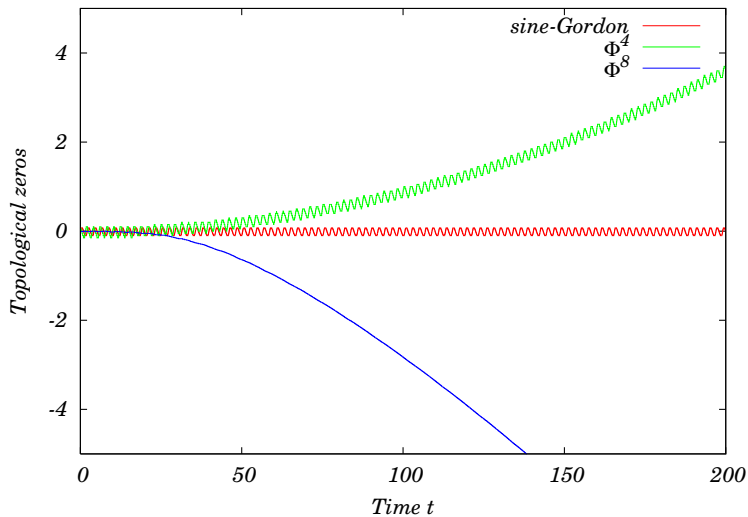}
\caption{Paths of the zeros of the kink in the sine-Gordon, $\phi^4$ and $\phi^8$ models.
The zeros of $\phi^8$ kink were scaled by 0.1.\label{f1.1}}
\end{figure}

It is quite illuminating to compare the prediction for the acceleration  of the
kink \eqref{eq:eq1} by solving numerically the nonlinear wave equation (\ref{eq:eq1}) for a few
common one dimensional field theories.
On Figure \ref{f1.1} we have depicted the trajectories of the (topological) zeros of various kink
solutions under the influence of an incoming radiation from $+\infty$ in three models:
in the sine-Gordon, $\phi^4$ and in the $\phi^8$ model where $U=\frac{1}{4}(\phi^2-1)^4$.
It is somewhat surprising that the kink has been pushed by the radiation pressure
only in the very last example, in agreement with Eq.\ (\ref{F1}) .
As it can be seen on Fig. \ref{f1.1} the time average of the acceleration of the kink in the
sG model is zero, the kink is steadily oscillating around its initial position.
Most remarkably the kink in the $\phi^4$ model
accelerates towards the source of radiation, and
it is this interesting effect that we interpret as {\sl negative radiation pressure}.
Taking into account other collective coordinates such as the shape mode,
would not substantially influence our main results, therefore we have chosen to ignore them.
Note that the acceleration of the kink in the $\phi^8$ model is noticeably larger
as compared to the $\phi^4$ one, for the same amplitude of the incoming wave ($A=0.14$).
In fact, while according to Eq.\ \eqref{F1} the acceleration of the kink is {\sl quadratic}
in the amplitude $A$, this is only true for the $\phi^8$ model.
The acceleration of the $\phi^4$ kink turns out to be proportional to $A^4$.
Motivated by the unexpected results shown on Figure \ref{f1.1} we shall investigate
the interaction of a kink with radiation in more detail in order to give an explanation of
the negative radiation pressure.
It is immediately clear that the derivation of the force obtained in Eq.\ (\ref{F1})
is only valid if its leading contribution comes from the linear approximation.
In the case when $|{\mathbf R}|\ll1$ the validity of this assumption is questionable
since then the higher order terms may contribute in an important way,
therefore the first order result in Eq.\ \eqref{F1} is not necessarily correct.
As a matter of fact it is rather well known that in both the $\phi^4$ and in the sG models
that the potentials \eqref{potentials} are reflectionless, \ie ${\mathbf R}=0$, for all frequencies.
Therefore in such models the dynamics of the kinks is determined by higher order terms.

\section{Computation of the force on the kink up to $\Or(A^4)$}\label{section:force4}

In this Section we shall outline the computation of the force
acting on the kink as well as its acceleration to higher orders of perturbation
theory. As we shall show below the next nontrivial correction to the force
turns out to be of $\Or(A^4)$. Fortunately one does not have to go
up to $4$th order in perturbation theory.

By expanding the equation of motion of the kink \eqref{eq:eq1}
in power series of the amplitude of the incoming wave, $A$, corresponding to \eqref{eq:post1},
the $n$th order solution of the equation of motion is determined by the inhomogeneous linear
equations
\begin{equation}\label{eq:nr1}
{\ddot{\xi}}^{(n)}+\opl \xii{n} = f^{(n)},
\end{equation}
where the source terms $f^{(n)}$ can be calculated from the lower order terms
in the perturbation series, see Appendix \ref{appendix:genpert} for details.
To define the solution of the perturbative equations \eqref{eq:nr1}
uniquely, we impose
in each order that there are no incoming waves from the left ($x<0$), as boundary conditions.
These correspond to the physical problem of an incoming wave from
the r.h.s.\ of the kink.
We remark that to avoid resonances in higher orders, the frequency, $\omega$,
has also to be expanded as
\begin{equation}
\omega=\omega^{(0)}+A\omega^{(1)}+A^2\omega^{(2)}+\cdots. \label{eq:czes}
\end{equation}

The total force acting on the system ``kink $+$ radiation'' inside the segment $[-L,L]$,
can be calculated to
higher orders in $A$ from the energy and momentum conservation laws
similarly to the leading order computation.
We compute the rate of change of the total momentum inside the box $[-L,L]$
to obtain the total force acting on the system, which after
averaging in time can be consistently
identified to leading order with the force {\sl acting on the kink} just as in the
calculation in the previous section.

Using $\phi(x,t)=\phi_s+\xi$ in Eq.\ (\ref{eq:ZasP}) and integrating
over the interval $[-L,L]$, one finds
\begin{equation}\label{eq:noether2}
\left.\partial_t P=-\frac{1}{2}\left(\dot{\xi}^2+\xi'^2-U''(\phi_{})\xi^2-
\frac{1}{3}U'''(\phi_{})\xi^3-\cdots\right)\right|^{L}_{-L}\,,
\end{equation}
where we have omitted terms which are exponentially small
for $L\gg1$. Choosing $L$ to be sufficiently large,
one can use the asymptotic form of the higher order solutions
to find the rate of change of the averaged momentum inside the segment $[-L,L]$
just as for the lowest order case.
Assuming that $\partial_t P=0$ to $\Or(A^2)$
in Eq.\ \eqref{eq:noether2} (reflectionless case)
one finds that all terms of order $\Or(A^3)$ also drop out,
i.e.\ there is no momentum flow to this order into the segment
(for more details see Appendix A).
Therefore the first non-trivial contribution to Eq.\ \eqref{eq:noether2} is of order
$\Or(A^4)$, implying that one would also need $\xi^{(3)}$,
i.e.\ one should compute up to 3rd order in perturbation theory.
This complicates considerably the problem, even if only the asymptotic forms of the $\xi^{(k)}$
are needed. For some computational details of the higher order calculations
we refer to Appendix A.
Remarkably though one can in fact eliminate the contributions coming from the 3rd order terms
from the momentum balance in the segment $[-L,L]$ by exploiting the law of energy conservation.
As it will be shown below all the information needed to calculate the force acting on the kink
is actually encoded in the asymptotic form of the time-dependent part of $\xi^{(2)}$.

Denoting by $\xi^{(n)}_m$ the $m$-th coefficient in the Fourier expansion of $\xi$ in the
$\Or (A^n)$ order, the $\Or(A^2)$ order solutions can be written as:
\begin{equation}\label{order2}
\xi^{(2)}=e^{2i\omega t}\xi^{(2)}_{+2}+\xi^{(2)}_0+e^{-2i\omega t}\xi^{(2)}_{-2}\,.
\end{equation}
We also note that for reflectionless potentials ${\xi^{(2)}_{+2}}^*=\xi^{(2)}_{-2}$.
The asymptotic form of the time-dependent part of the 2nd order term, $\xi^{(2)}$,
has the form
\begin{equation}\label{eq:xi22_gen}
\xi^{(2)}_{+2}(x\rightarrow\pm \infty)=\frac{U'''(\phi_{vac})}{24U''(\phi_{vac})}\eta_{+q}^2
+\alpha_{22,\pm k}(q)\eta_{\mp k}\,,
\end{equation}
where
\begin{equation}\label{alpha22}
\alpha_{22,k}(q)=-\frac{1}{8W}\int_{-\infty}^\infty\!\!\!dx'\;\eta_{k}\eta_q^2U'''(\phi_s)\,,
\end{equation}
$k=\sqrt{4\omega^2-U''(\phi_{vac})}$ is the wave number corresponding
to $2\omega$, and $W=\eta_k\eta_{-k}'-\eta_k'\eta_{-k}=-2ik$ is the Wronskian.
In a way $\alpha_{22,k}(q)$ encodes the reflection and transition coefficients due to the non-linear
effects.
The details of the computation of Eq.\ \eqref{eq:noether2} up to $\Or(A^4)$ can be found
in Appendix A, leading to the result
\begin{equation}\label{eq:ZasPP}
\langle\pa_t P\rangle_T=F^{(4)}=-A^4\left[2k^2\left(|\alpha^2_{22,+k}|-|\alpha^2_{22,-k}|\right)-
2q^2\Re(\alpha_{31,-q})\right]\,,
\end{equation}
where the pertinent contribution from the 3rd order terms is encoded in a single coefficient
$\alpha_{31,-q}$, defined in Eq.\ (\ref{eq:xi3asy}).
In Eq.\ \eqref{eq:ZasPP} we have identified the time average
of the overall momentum, $\langle\pa_t P\rangle_T$, flowing into the segment $[-L,L]$,
with the force, $F$, exerted by the incoming radiation on the kink.

In the case of reflectionless potentials the first non-vanishing contribution
to the acceleration comes from the $\Or(A^4)$ terms. Considering for simplicity
such theories where the kink is spatially antisymmetric (this includes both ``$\phi^4$'' and the sG models)
a straightforward computation yields:
\begin{equation}\label{eq:accel0_norm}
m_s a^{(4)}=-{A^4}(\ddot{\xi}^{(4)}_0|\eta_t)=
\Re( U'''(\phi_s)[\xi^{(3)}_1\eta_{-q}+\xi^{(2)}_2\xi^{(2)}_{-2}]+
U^{(\rm iv)}(\phi_s)\xi_2^{(2)}\eta_{-q}^2/4|\eta_t)\,,
\end{equation}
i.e.\ it is sufficient to compute the second and third order solutions
(see Appendix \ref{appendix:genpert} for more details).
By a direct computation we have checked that
with the definition of the force in Eq.\ \eqref{eq:ZasPP} acting on the kink
within our perturbative framework, Newton's law
\be\label{Newton}
F^{(4)}=m_sa^{(4)}\,,
\ee
still holds, at least up to the fourth order $\Or(A^4)$.
In our view this result lends strong support to identify
the time average of the momentum flow in the segment $[-L,L]$
with the force acting on the kink to leading order in PT.

We give here a derivation of Newton's law, which also indicates the
limits of its validity.
Using the following simple identity
\be
\ddot\xi\phi_s'=\pa_t(\dot\phi\phi'-\dot\xi\xi')=-\pa_t({\cal P}+\dot\xi\xi')\,,
\ee
by integrating over a segment $[-L,L]$ one easily obtains the relation
\be\label{moment1}
\int\limits_{-L}^L dx \ddot\xi\eta_t=-\dot P_L-\int\limits_{-L}^L dx \pa_t(\dot\xi\xi')\,,
\quad \dot{ P}_L=\left.\dot{ P}\right\vert^{-L}_L\,
\ee
where $\dot P_L$ denotes the momentum change inside the segment $[-L,L]$.
We shall approximate the lhs of Eq.\ \eqref{moment1} simply by $(\ddot\xi|\eta_t):=-m_sa$
since the difference between them is exponentially small in $L$.
After averaging in time we obtain
\begin{equation}\label{newton1}
F_L=m_sa-\int\limits_{-L}^L dx\left\langle(\pa_t(\dot{\xi}\xi')\right\rangle_T\,,
\end{equation}
where $F_L$ denotes the total force acting on the box.
Now in perturbation theory the solution $\xi$ can be decomposed as
\be\label{xi-decomp}
\xi=\xi_p(t,x)+\xi_0(x)-\frac{1}{2}at^2\xi_t(x,t)
\ee
where $\xi_p(t,x)$ is periodic in time and
it is at least of order $\Or(A)$,
$\xi_t$ corresponds to the ``accelerating part'' (with initially constant acceleration),
and $\xi_0(x)$ is the time independent piece.
This holds to order $\Or(A^2)$ when $a^{(2)}\ne0$, resp.\ to order $\Or(A^4)$
when $a^{(2)}=0$:
\be\label{xi-decomp4}
\xi_p=A\xi^{(1)}+A^2\xi^{(2)}+\Or(A^3)\quad \xi_0(x)=A^2\xi^{(2)}_0(x)+\Or(A^4)\quad
\xi_t=A^n\eta_t(x)+\Or(A^{n+2})\,,
\ee
where $n=2$ if $a^{(2)}\ne0$, and $n=4$ if $a^{(2)}=0$.
The leading order correction is $\propto aA^{n+1}$, which
depends on time averaging and indicates the limits of the validity of our simple minded
approach.

Let us note here that the mass of the kink gets renormalized due to its interaction
with the radiation field.
A standard calculation for the first correction to the kink mass is given as
$m^*=m_s+A^2\delta m^{(2)}\cdots$, i.e.\ to lowest nontrivial order
$\delta m$ is proportional to $A^2$, therefore to $\Or(A^4)$ it does not show up
in Eq.\ \eqref{Newton}. Nevertheless the numerical simulations (see Section \ref{section:numer})
indicate that the kink has an effective mass, whose value is quite close to
the renormalized mass.

The computation of the rate of change of the energy inside the segment is completely
analogous to the previous momentum balance calculation, we find:
\begin{equation}\label{eq:37}
\langle\partial_t E\rangle=-A^4\left[4\omega k\left(|\alpha^2_{22,k}|+|\alpha^2_{22,-k}|\right)+
2\omega q\Re(\alpha_{31,-q})\right]\,.
\end{equation}
Assuming that after averaging, at least for some initial time the kink
can be considered as a rigidly accelerating particle, if it
was initially at rest, i.e.\ $v(t=0)=0$ then obviously
\begin{equation}\label{deltaE}
\langle\partial_t E\rangle|_{t=0} = m v \dot v|_{t=0} = 0\,.
\end{equation}
This equation together with eq.\ (\ref{eq:37}) can be now used to eliminate
the coefficient $\alpha_{31,-q}$ from Eq.\ (\ref{eq:ZasPP}), and then
one obtains a remarkably simple formula determining
the force acting on the kink:
\begin{equation}\label{eq:sila}
F^{(4)}=2A^4k\left[(k-2q)|\alpha^2_{22,-k}|-(k+2q)|\alpha^2_{22,k}|\right]\,.
\end{equation}
By a direct computation of the energy we have verified that $\langle\partial_t E\rangle|_{t=0}$
is indeed zero up to $\Or(A^6)$, which shows the validity of Eq.\ \eqref{deltaE}.

To conclude this Section, we have calculated the force exerted by
an incoming wave on the kink to the first non-trivial
order in perturbation theory in the class of models where the linearization
around the kink yields a reflectionless potential.
It is important to emphasize that for the class of models
where the effective potential is reflectionless,
the force exerted on the kink turns out to be proportional to $F\sim\Or(A^4)$.
This is to be contrasted to more generic models where the effective potential
is reflective, in which case $F\sim\Or(A^2)$.

\section{Negative radiation pressure in the $\phi^4$ model}\label{section:phi4}
In this Section we apply the previously obtained general results
to compute explicitly the force exerted on the kink
by an incoming wave in the $\phi^4$ and in the SG models.

In the $\phi^4$ model the full nonlinear equation for the ``radiation" $\xi$ is:
\begin{equation}\label{npeq1}
\ddot{\xi}+\opL\xi+6\phi_s\xi^2+2\xi^3=0\,.
\end{equation}

The first order solution $\xi^{(1)}$ in Eq.\ \eqref{xi1}
can be explicitly given both in the $\phi^4$ and in the sG models:
\begin{equation}\label{eq:etaqp4}
\eta_q^{\phi^4}=\frac{3\tanh^2x-1-q^2-3i q\tanh x}{\sqrt{(q^2+1)(q^2+4)}}e^{i qx}\,,\quad
{\rm where}\quad q^2+4=\omega^2 \,,
\end{equation}
and
\begin{equation}\label{eq:etaqsG}
\eta_q^{\rm sG}=\frac{i q -\tanh x}{\sqrt{q^2+1}}e^{i qx}\,,\quad
{\rm where}\quad q^2+1=\omega^2 \,.
\end{equation}
The 2nd order solution ($\Or(A^2)$) given by Eq.\ \eqref{order2} contains the zero frequency
term, $\xi^{(2)}_0$, which in general depends both on $t$ and $x$.
In the present case it is consistent to assume that $\xi^{(2)}_0$ is time-independent (see
in Appendix \ref{appendix:genpert} for a proof), moreover it can be
written explicitly.
The 2nd order ``transition and reflection'' coefficients, $\alpha_{22, k}(q)$ in Eq.\ \eqref{alpha22}
can also be calculated analytically, the result is:
\begin{equation}\label{alpha22-fi4}
\alpha_{22, k}(q)
= -\frac{3}{2}\pi\frac{q^2+4}{q^2+1}\sqrt{\frac{q^2+4}{k^2+1}}\;
\frac{1}{k\sinh\left(\frac{2q+k}{2}\pi\right)}\,,
\end{equation}
where $q^2=\omega^2-4$, $k^2=4(\omega^2-1)$.
Using Eq.\ \eqref{alpha22-fi4} it is now easy to find the averaged force \eqref{eq:sila}
exerted on the $\phi^4$ kink by an incident wave of frequency $\omega$, we obtain
\begin{equation}\label{force}
F^{(4)}=\frac{9\pi^2A^4\omega^6}{k(4\omega^2-3)(\omega^2-3)^2}\left[ \frac{\omega_-}{\sinh^2\pi\omega_-}-
\frac{\omega_+}{\sinh^2\pi\omega_+}\right]\quad{\rm with}\quad  \omega_\pm:=\sqrt{\omega^2-1}\pm \sqrt{\omega^2-4}\,.
\end{equation}
Introducing $f^{(4)}=F^{(4)}/A^4$, the behaviour of the function
$f^{(4)}(\omega)$ for $\omega\to2$ (i.e.\ small values of $q$) is given as
$f^{(4)}(\omega\to2)\approx0.3749\sqrt{\omega-2}$ while $f^{(4)}(\omega)\to3/4$ for
large values of $\omega$.
Quite interestingly the force acting on the kink
is {\sl positive}, therefore it accelerates towards the source of radiation.
This is the effect we refer to as {\sl negative radiation pressure}.
The origin of the negative radiation pressure can be understood by noticing that
for all frequencies, $\omega$, the amplitude of the (non-linearly) reflected wave
$|\alpha_{22,+k}|$ is smaller
than the amplitude of the transient wave $|\alpha_{22,-k}|$, i.e.\
$\alpha^2_{22,k}\ll\alpha^2_{22,-k}$.
In first order perturbation theory, such an effect would not be possible because of the identity
(\ref{cons1}) expressing energy conservation at the linear level.

The surprising effect of negative radiation pressure on the kink
exists only because of the presence of nonlinearities.
In the linear approximation the kink is transparent to the incident wave, therefore
it does not accelerate.
Because of the nonlinear terms, part of the energy of the incoming wave is transformed
into a wave whose frequency is twice of the original one.
This double frequency wave has a larger ratio of momentum to energy density
than the incident wave with smaller frequency, hence it
carries more momentum than the originally incident one.
This way a surplus of momentum is created behind the kink,
which pushes it towards the direction of the incoming wave.
The above is of course only an intuitive explanation of the negative radiation pressure
on the kink.
The effect of negative radiation pressure has been clearly observed in our numerical simulations
of the $\phi^4$ model (see the Section "Numerical simulation").

\begin{figure}
\PutFigure{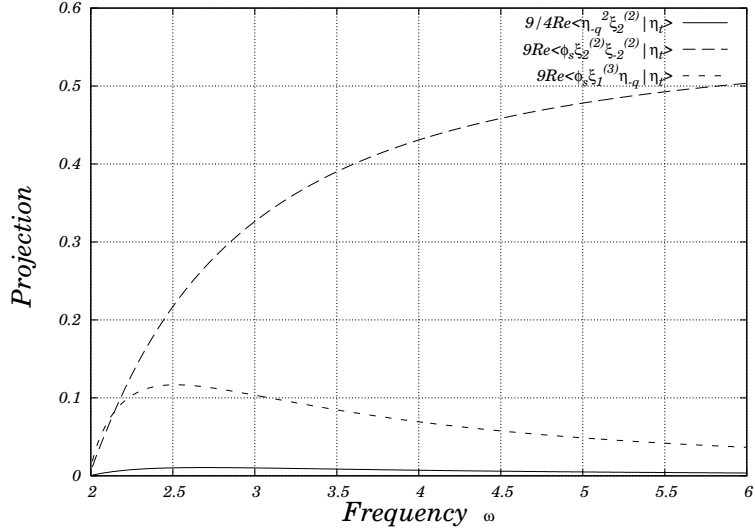}
\caption{Three projections obtained from the second and third order (needed for
\ref{eq:accel0_norm}). \label{f0}}
\end{figure}

Next we compute the acceleration of the $\phi^4$ kink using
Eq.\ \eqref{eq:accel0_norm} derived in the previous Section.
We need to compute in fact the projection of
the second and third order solutions on the translational mode, $\eta_t$.
We have computed them numerically by two different methods,
using the integral representation based on the explicitly known Green's function
and also by direct numerical integration of the corresponding equations \eqref{eq:nr1}.
We show separately the three projections in (\ref{eq:accel0_norm}) on Figure \ref{f0}
(divided by $(\eta_t|\eta_t)=4/3)$.
It is worthwhile to point out that all the three projections are positive.

Let us now turn to the sine-Gordon kink, and evaluate the force of order $\Or(A^4)$.
In the case of the sine-Gordon model our numerical results did not show any net radiation pressure,
the sG kink was oscillating around its initial position (c.f.\ Fig\ref{f1.1}).
Interestingly in the sG model the second order waveform can be calculated in closed form:
\begin{equation}
  \label{eq:so}
  \xi^{(2)}_{+2}={\frac {i q+\tanh \left( x \right) }
{ 16\left( 1+{q}^{2} \right) \cosh \left( x \right) }}\,e^{2i qx}\,,
\end{equation}
from which one can immediately see that the coefficients
$\alpha_{22,\pm k}$  determined by the asymptotic behaviour of $\xi^{(2)}$ in
Eq.\ \eqref{eq:xi22_gen}
are zero. This implies that the radiation exerts no force at all on sG kink at least up to
this order.
Clearly this interesting fact should be related to the special feature of the sG model, namely its
integrability.
As a matter of fact there is a remarkable analytic solution corresponding
to the nonlinear superposition of a kink with a travelling (cnoidal) wave in the sG model
obtained by H.~J.~Shin \cite{Shin} using the Darboux transformation method.
In Appendix \ref{appendix:shin}
we give a short review of Shin's solution which is somewhat complicated,
and we demonstrate that one can expand it in a
parameter which can be identified with the asymptotic amplitude of the incoming wave.
This way we have verified that
both the first and the second order solutions obtained by our perturbative
calculations agree perfectly with the Taylor expansion of the analytical
solution.
This comparison has also served as a test of the validity of our perturbative method.

\section{Numerical simulation}\label{section:numer}
In the present Section we outline the numerical method used to solve the
nonlinear wave equation Eq.\ (\ref{eq:eq1}) describing the interaction of
a kink with an incident wave and present the results in form of figures and tables.

We have discretized equation (\ref{eq:eq1}) in the spatial variable, $x$ as
$\phi(nh,t):=\phi_n(t)$.
The second derivative was approximated using the following five point scheme:
\begin{equation}
\phi''\equiv D\phi_n=\frac{1}{12h^2}\left(-\phi_{n-2}+16\phi_{n-1}-30\phi_n
+16\phi_{n+1}-\phi_{n+2}\right)+{\cal O}(h^4)\,.
\end{equation}
This way Eq.\ (\ref{eq:eq1}) reduced to a system of ordinary differential equations:
\begin{equation}
\ddot{\phi}_n=D\phi_n-U'(\phi_n)\,.
\end{equation}
We have simply put this coupled infinite system into a finite box of size $2L$,
which was then solved using a standard fourth order Runge-Kutta method.

Our initial conditions have been chosen to correspond to a kink together
with a first order travelling wave:
\begin{eqnarray}
\phi(x,t=0)&=&\phi_s(x)+\frac{1}{2}A\eta_q(x)+c.c.\,,\\
\dot{\phi}(x,t=0)&=&\frac{1}{2}i\omega A\eta_q(x)+c.c.\,,
\end{eqnarray}
and we have fixed the boundary values of $\phi(x,t)$ at $x=\pm L$ as
\be
\phi(x=\pm L,t)=\pm 1\,.
\ee
The evolution time of the system was restricted to be smaller than $L$, to avoid
the unphysical influence of the reflected waves from the artificial boundaries at $x=\pm L$
on the kink's motion.
The position of a static kink can be quite unambiguously identified
by the location of its zero which coincides with the maximum of its energy density.
It is less clear how to define the position of an interacting kink. In our case
one has to separate first the field of the kink from that of the radiation,
which can already be problematic and the position of the kink is not very well defined,
in general the maximum of the energy density and the zero of $\phi(x,t)$ do not coincide.
For small enough amplitudes the kink is only slightly perturbed and therefore
its topological zero is still a rather satisfactory definition as the position of the kink
and we have used this definition in our work.

We have plotted the position of the zero of $\phi(x,t)$ as a function of time for the frequency
$\omega=3.0$ and for the amplitude of the wave $A=0.12$ on Figure 3.
\begin{figure}
\begin{center}
\PutFigure{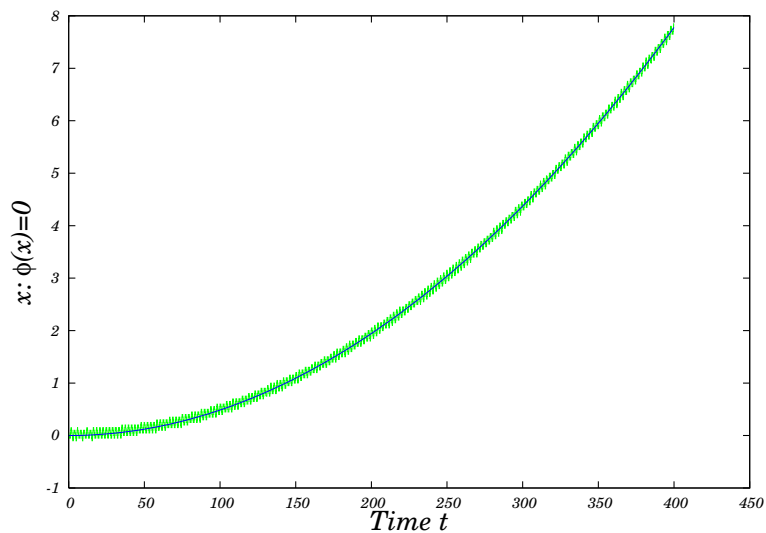}
\end{center}
\caption{The position of the zero of $\phi$ as a function of time for $A=0.12$, $\omega=3.0$.}\label{freq}
\end{figure}
On this figure one can clearly see that the trajectory of the zero of $\phi(x,t)$
is quite close to a parabola, corresponding to the expected non-relativistic acceleration of
the kink.
A  numerical fit confirms that a parabola of the form $at^2/2$ to the trajectory
is good a approximation indeed.
The fitted value of the acceleration in this case was $a_{\rm num}=9.72\cdot10^{-5}$ which is
not that far
from the result of our analytical calculations in Eq.\ \eqref{eq:accel0_norm} giving
$a_{\rm theor}=9.02\cdot10^{-5}$.
Taking into account that in the analytical calculation only the leading terms
have been used, this agreement appears to us satisfying.
Next we have checked if the measured acceleration is indeed $\Or (A^4)$ as predicted by the
leading order
perturbative result \eqref{eq:accel0_norm}.
On Figure \ref{f3.4} we have plotted the fitted acceleration for $\omega=3.0$ divided by $A^4$
for the amplitudes of incoming wave varying between $0.1\leq A\leq 0.3$.
As one can see for $0.1\leq A\leq 0.22$ the curve is close to being flat
implying that the dominant term is indeed proportional to $A^4$.
This proportionality breaks down when the value of the amplitude increases
to about $A\approx0.24$. For $A=0.25$ even the sign of the fitted acceleration changes.
\begin{figure}
\PutFigure{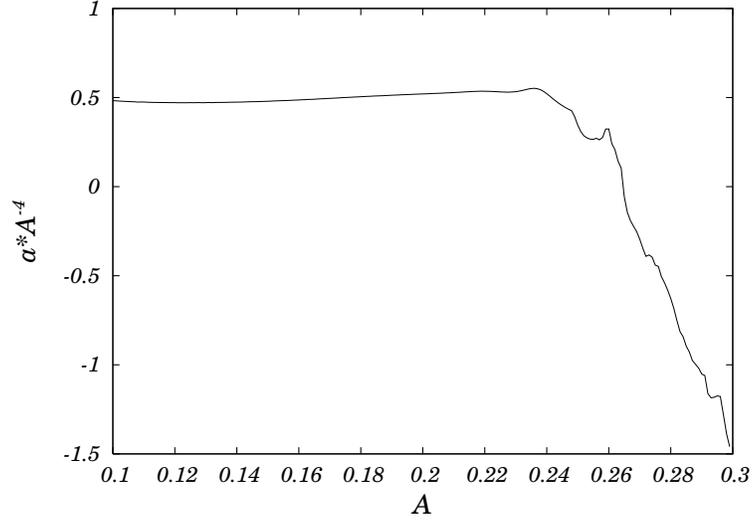}
\caption{Fitted acceleration divided by  $A^4$ for $\omega=3.0$ as a function of $A$. \label{f3.4}}
\end{figure}
In Table I we have compared the numerically obtained values of the acceleration
to the theoretical ones for a range of amplitudes.
\begin{center}
\begin{table}
\begin{tabular}{|c|c|c|}
\hline
Amplitude $A$&Fitted acceleration &Theoretical value\\
\hline
0.10  &  0.0000482 &  0.00004354\\
0.12  &  0.0000977 &  0.00009029\\
0.16  &  0.0003188 &  0.00028537\\
0.18  &  0.0005296 &  0.00045710\\
0.20  &  0.0008325 &  0.00069670\\
0.22  &  0.0012541 &  0.00102005\\
0.24  &  0.0017300 &  0.00144469\\
0.26  &  0.0014806 &  0.00198986\\
\hline
\hline
\end{tabular}
\caption{Fitted and theoretical values of the acceleration for $\omega=3.0$}\label{tab:tab2}
\end{table}
\end{center}
From Table I one can see that the agreement between the calculated
and the fitted values of the acceleration is reasonably good up to values of $A<0.22$.
These results confirm that for amplitudes of the incoming wave
in the range $0.1\leq A\leq0.22$ the kink accelerates non-relativistically, and
also that its acceleration scales as $A^4$.
Next we exhibit the numerically obtained acceleration in function of
the frequency $\omega$ on Figure \ref{f3.2}, together with the theoretical curve,
and some results are given in Table II.
\begin{figure}
\PutFigure{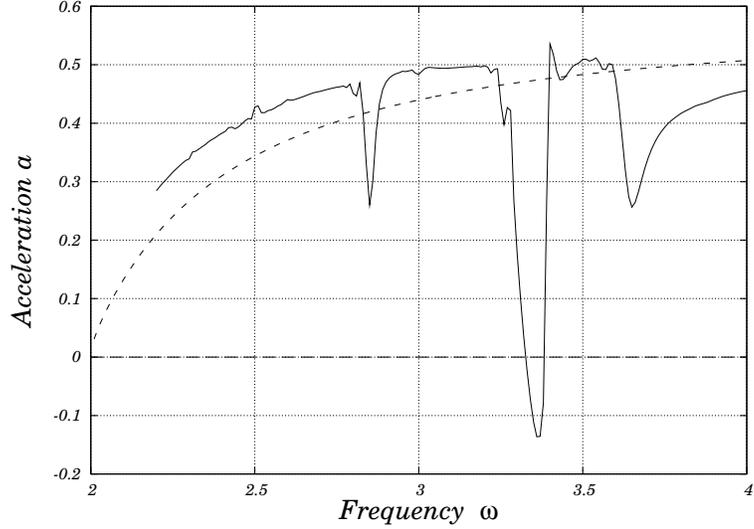}
\caption{Fitted acceleration divided by $A^4$ for $A=0.16$ (solid line)
and the acceleration calculated analytically (dashed line). \label{f3.2}}
\end{figure}
\begin{center}
\begin{table}
\begin{tabular}{|c|c|c|}
\hline
Frequency $\omega$&Fitted acceleration $aA^{-4}$&Theoretical value\\
\hline
2.50 &   0.4285& 0.339743\\
2.70 &   0.4472& 0.388489\\
2.90 &   0.4397& 0.421959\\
3.10 &   0.4865& 0.446130\\
3.30 &   0.3510& 0.464260\\
3.50 &   0.4761& 0.478267\\
3.70 &   0.3310& 0.489349\\
3.90 &   0.4322& 0.498290\\
4.30 &   0.4672& 0.505624\\
\hline
\hline
\end{tabular}
\caption{Fitted acceleration divided by  $A^4$ for $A=0.16$}\label{tab:tab1}
\end{table}
\end{center}
The first thing one might notice on Figure \ref{f3.2} is the presence of three resonance-like
structures completely absent from the theoretical curve which is a monotonously increasing
function of
$\omega$.
The largest resonance is not very far from $2\omega_d\approx3.46$ which indicates that
it is likely to be related to the coupling between internal (or shape) mode of the kink and
radiation.
A plausible explanation of the important change in the acceleration at frequencies when
the shape mode couples strongly to the incoming wave is the following.
At such a ``resonance'' frequency the shape mode accumulates a substantial
amount of energy, which is then radiated symmetrically in both directions.
Far from the kink this produces the same effect as a reflected wave, thus,
at such resonant frequencies the kink is not transparent.
\begin{figure}
\PutFigure{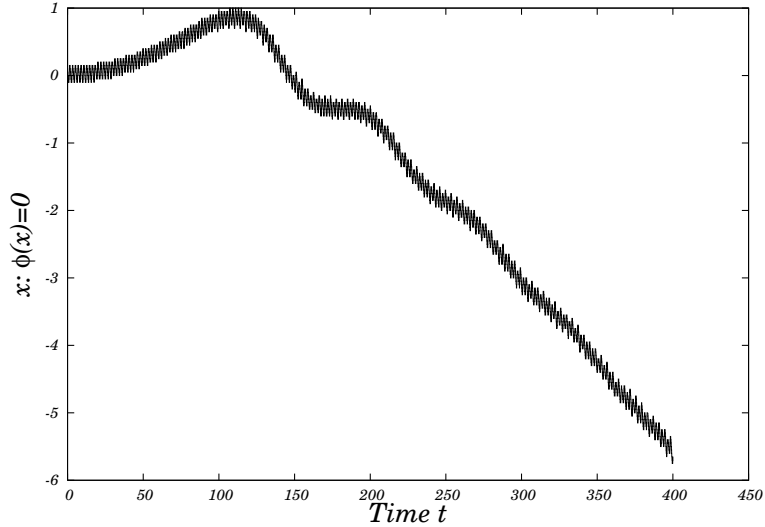}
\caption{The position of the zero of $\phi$ as a function of time for $\omega=3.38$ and $A=0.14$.}\label{f3.3}
\end{figure}
On Figure \ref{f3.3} the path of the kink is plotted for a value of the frequency near the
resonance. As one can see the motion of the kink is somewhat irregular there.
All in all the discrepancy between the results
of the perturbative computations to leading nontrivial order
and those of the numerical simulation does not exceed
10\% for a large range of the frequencies with the exception of three resonances.
It seems to us that in view this agreement is satisfactory in view of the
approximations used.

\begin{figure}
\PutFigure{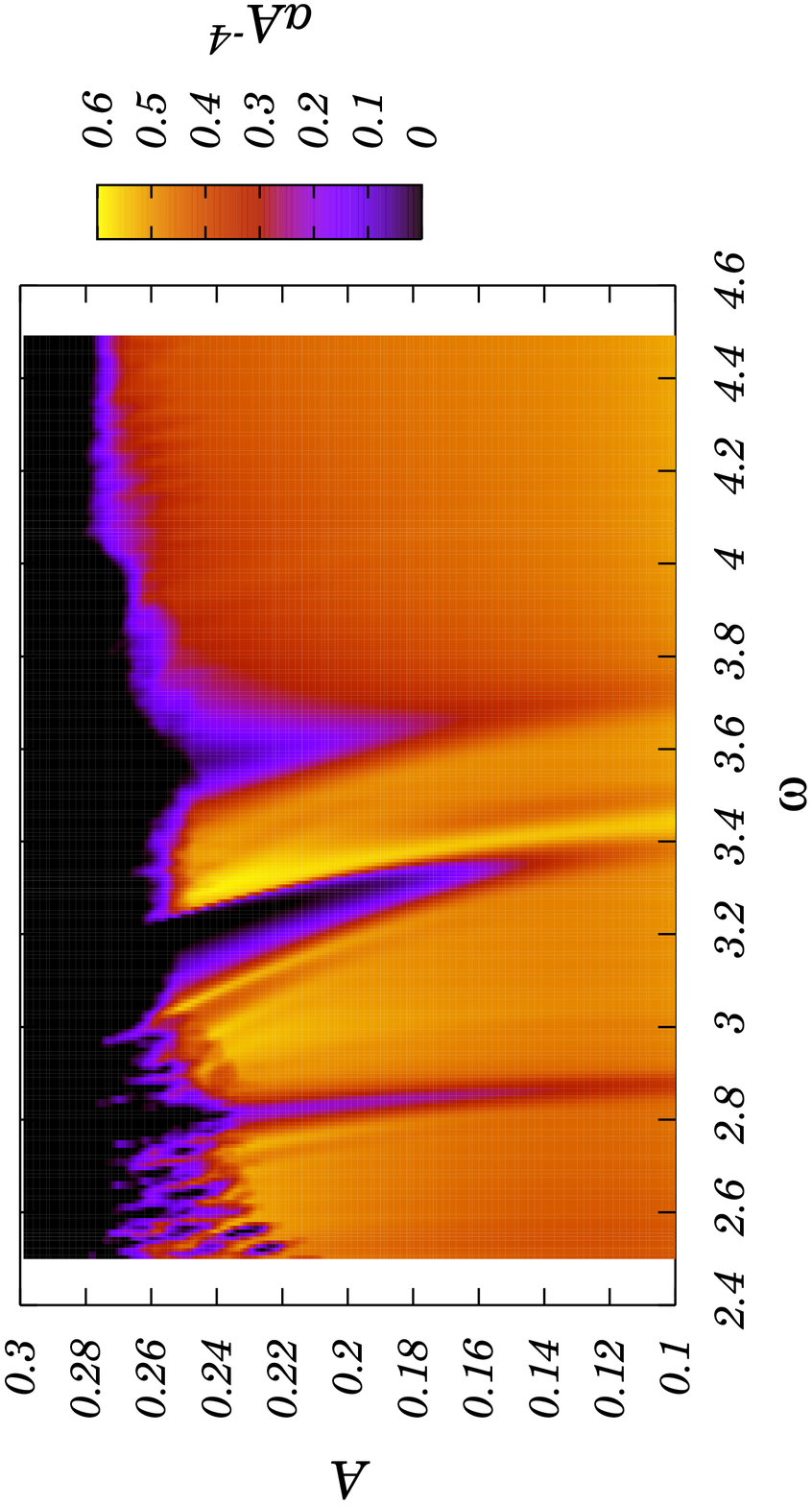}
\caption{Acceleration $aA^{-4}$ as a function of amplitude and frequency of incoming wave.\label{scan}}
\end{figure}
\begin{figure}
\PutFigure{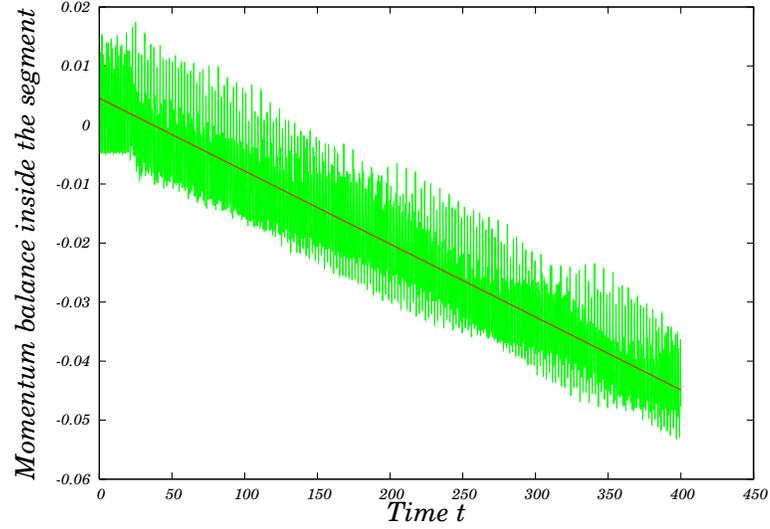}
\caption{Momentum balance inside the segment [-20,20] for  $\omega=3.0$ and \mbox{$A=0.12$}.
The slope of the fitted straight line (as a sort of averaging) is \mbox{$-1.24\cdot10^{-5}$}, corresponding
to an effective mass $m^*=0.954 m$.
\label{f4}
\vspace*{-0mm}}
\end{figure}

As it can be seen from Table \ref{tab:tab2}, the numerically found acceleration is
systematically larger than the leading order theoretical one.
Clearly higher order effects could play a role here, and
the simplest one to be taken into account is the renormalization of the kink mass
due to the radiation field.
The lowest order $\Or(A^2)$ contribution to the mass, $\delta m^{(2)}$ is negative
for all frequencies, i.e.\  the effective mass $m^*=m+A^2\delta m^{(2)}<m$ which goes into the
right direction. It is quite difficult to obtain a sufficiently precise numerical value
for the effective mass, nevertheless we have obtained some indicative results.
The numerically computed momentum balance (force) inside a box [-20,20] is
presented on Figure \ref{f4}.
Dividing this force by the measured acceleration we obtain the effective mass, $m^*$.
The mass obtained in this case is $m^*_{\rm\scriptscriptstyle num}=0.954m_s$.
An analytical calculation
yields $-\delta m^{(2)}=3 A^2(\omega^2-2)/(\omega^2-3)$, which gives
$m^*_{\rm\scriptscriptstyle theor}=0.962m_s$, so there is a reasonable degree
of agreement between the two.
In conclusion the numerical results show that our perturbative calculations
are quite reliable for amplitudes $A<0.2$ and for
frequencies far from resonance points.
Finally on Figure \ref{scan} the behaviour of the acceleration of the
$\phi^4$ kink for a range of amplitudes and frequencies
of the incoming wave is depicted.

\section{Stability of the effect under perturbations}\label{section:p4p}
In this Section we shall demonstrate that the effect of negative
radiation pressure has a certain degree of robustness with respect to perturbations
of the original $\phi^4$ model. This fact makes the effect, which is in itself interesting,
much more relevant for physical applications.
At first sight it is not so obvious that this effect could survive a small perturbation
of the model at all, since a generic perturbation, no matter how small it be, destroys the
reflectionless nature of the potential in Eq.\ \eqref{eq4}. This way
a first order perturbative contribution (in the amplitude $A$) is generated.
Therefore the leading term in the expression for the force
changes under the influence of a generic perturbation
from being of order $\Or(A^4)$ in Eq.\ (\ref{eq:sila})
to $\Or(A^2)$ as in Eq.\ (\ref{F1}). We shall show that although a small, but generic perturbation
of the $\phi^4$ model changes the leading term for the force to being of order
$\Or(A^2)$ indeed, for an important frequency range still the $\Or(A^4)$ will dominate
if the amplitude $A>A_{\rm crit}$.
We shall consider a concrete example of perturbation which illustrates
that the critical amplitude $A_{\rm crit}$ can turn out to be small
and that the negative radiation pressure stays practically unaffected.

To start with we shall consider a generic perturbation of the field equation
Eq.\ \eqref{eq:eq1} of the form:
\begin{equation}\label{pert-eq1}
\ddot{\phi}-\phi''+U'(\phi)+\epsilon\delta U'(\phi)=0\,,
\end{equation}
where $\epsilon$ is a small parameter and $\delta U'(\phi)$ is the perturbation.
We look for the solution of the perturbed equation \eqref{pert-eq1} again expanded
in a power series in the amplitude of the incoming wave:
\begin{equation}
\phi=\phi_s(x)+A\xi^{(1)}(x,t)+\cdots\,,
\end{equation}
where $\phi_s(x)$ denotes the static kink solution of the perturbed field
equation \eqref{pert-eq1} and $\xi^{(1)}(x,t)$ satisfies the field equation
linearized around $\phi_s(x)$,
\be\label{pert-xi1}
\ddot{\xi}^{(1)}-{\xi^{(1)}}''+U''(\phi_s)\xi^{(1)}+\epsilon\delta U''(\phi_s)\xi^{(1)}=0\,,
\ee
Now we also expand the solution of Eq.\ \eqref{pert-eq1} in $\epsilon$ determining
the perturbation of the original theory:
\begin{equation}
\phi_s(x)=\phi^{(0)}_s+\epsilon\phi^{(1)}_s+\cdots\,,\quad
\xi^{(1)}=\xi^{(10)}+\epsilon\xi^{(11)}+\cdots\,,
\end{equation}
where $\phi^{(0)}_s(x)$ is the static kink of the unperturbed equation \eqref{eq:eq1},
and $\xi^{(10)}(x,t)$ is a solution of the linearization of the unperturbed
field equation around $\phi^{(0)}_s(x)$, i.e.\  $\ddot{\xi}^{(10)}+\opl\xi^{(10)}=0$.
The equations for the first order corrections in the perturbative
parameter $\epsilon$ are:
\begin{eqnarray}\label{pert-xi1eq}
-{\phi^{(1)}_s}''+U''\left(\phi^{(0)}_s\right)\phi^{(1)}_s+
\delta U'\left(\phi^{(0)}_s\right)&=&0\,,\\
\ddot{\xi}^{(11)}-{\xi^{(11)}}''+U''\left(\phi^{(0)}_s\right)\xi^{(11)}
+U'''\left(\phi^{(0)}_s\right)\phi^{(1)}_s\xi^{(10)}
+\delta U''\left(\phi^{(0)}_s\right)\xi^{(10)}&=&0\,.
\end{eqnarray}
Let us recall that $\xi^{(10)}(x,t)=\frac{1}{2}e^{i\omega t}\eta^{(0)}_q+c.c.$,
and look for the solution as $\xi^{(11)}=\frac{1}{2}e^{i\omega t}\eta^{(1)}_q+c.c.$, where
$\eta^{(1)}_q(x)$ satisfies the following equation:
\begin{equation}\label{eta1}
\left(\opL-\omega^2\right)\eta_q^{(1)}+
\left[\delta U''\left(\phi^{(0)}_s\right)+U'''\left(\phi^{(0)}_s\right)\phi^{(1)}_s
\right]\eta_q^{(0)}=0\,.
\end{equation}
Since \eqref{eta1} is an inhomogeneous equation of the form of \eqref{eq:gen1}
its general solution can be obtained from Eq.\ \eqref{eq:gensol}.
In order to obtain the force we need the reflection coefficient, ${\mathbf R}$,
therefore it is sufficient to compute the asymptotic behaviour
of $\eta_q^{(1)}$ for large $|x|$, which can be found to given as:
\begin{eqnarray}\label{beta_R}
\eta^{(1)}_q(x\rightarrow+\infty)&=&-\frac{\eta_{-q}^{(0)}}{W}
\int_{-\infty}^\infty dx
\left[\delta U''\left(\phi^{(0)}_s\right)+U'''\left(\phi^{(0)}_s\right)\phi^{(1)}_s
\right]{\eta_q^{(0)}}^2:=
\beta_{R}^{(1)}\eta_{-q}^{(0)}\,,\\
\eta^{(1)}_q(x\rightarrow-\infty)&=&-\frac{\eta_{+q}^{(0)}}{W}
\int_{-\infty}^\infty dx{
\left[\delta U''\left(\phi^{(0)}_s\right)+U'''\left(\phi^{(0)}_s\right)\phi^{(1)}_s\right]
\eta_q^{(0)}}\eta_{-q}^{(0)}
:=\beta_{T}^{(1)}\eta_{q}^{(0)}\,.
\end{eqnarray}
The expressions $\epsilon\beta_{R}^{(1)}$ and $\epsilon\beta_{T}^{(1)}$ are the
first non-trivial corrections to the reflection (${\mathbf R}$) and transition (${\mathbf T}$) coefficients.
Recall that in the reflectionless case, such as in the $\phi^4$ model ${\mathbf R}=0$
and $|{\mathbf T}|=1$.
From Eq.\ \eqref{beta_R}
it immediately follows that in the perturbed field equation,
\eqref{pert-eq1} the kink is not transparent anymore.
The leading contribution to the force is then determined by the
first order term linear in the amplitude $A$.
As found in Eq.\ \eqref{F1} the dominant part of the force acting on the kink, $F^{(2)}$,
is proportional to the square of the reflection coefficient, i.e.
\begin{equation}
F^{(2)}:= A^2f^{(2)}=-q^2|{\mathbf R}^2|A^2\approx -q^2\epsilon^2|\beta_{R}^{(1)}|^2A^2\,.
\end{equation}
The above equation holds if $|\epsilon\beta_{R}^{(1)}|\,, |\epsilon\beta_{R}^{(1)}|\ll1$,
which is true for sufficiently small values of $\epsilon$
and for a certain range of $q$.
The first perturbative correction in $\epsilon$ to the force of order ${\cal O}(A^4)$
in the amplitude, $F^{(4)}\equiv A^4f^{(4)}$ (c.f.\ Eq.\ (\ref{eq:sila}))
will be at least of order ${\cal O}(A^4\epsilon^2)$ therefore this term can be neglected
in the following. The leading contribution to the force acting on the
kink due to a perturbation of a theory with reflectionless potential is given as
\be\label{Fcrit}
F=A^2(-q^2\epsilon^2|\beta_{R}^{(1)}|^2+A^2f^{(4)})\,.
\ee
Assuming that $f^{(4)}>0$ (i.e.\ that the radiation pressure is negative in the model)
it follows from Eq.\ \eqref{Fcrit} that the amplitude of the incoming wave
must be larger than a critical value, $A>A_{\rm crit}$ for a fixed value of $\epsilon$
to ensure $F>0$, i.e.\ that the effect of the negative radiation pressure be present.
The value of the critical amplitude is determined
by the condition $F^{(2)}+F^{(4)}=0$, leading to
\begin{equation}\label{A-crit}
A_{\rm crit}=\frac{\epsilon q|\beta_{R}^{(1)}|}{\sqrt{f^{(4)}}}.
\end{equation}
Clearly the result in \eqref{A-crit} is meaningful in our perturbative
framework, only if $A_{\rm crit}\ll1$.

We now apply the above general results to the $\phi^4$ theory
(where $U'(\phi) = 2\phi(\phi^2-1)$). We have chosen the following perturbation
for $\delta U'(\phi)$:
\be\label{fi6}
\delta U'(\phi)=\phi(\phi^2-1)^2\,.
\ee
Recall that $\phi^{(0)}_s(x)$ is nothing but the static kink
in the unperturbed $\phi^4$ theory, so
\begin{equation}
\phi^{(0)}_s(x)=\tanh x\,.
\end{equation}
The first order correction in $\epsilon$ to the static kink is determined by Eq.\
\eqref{pert-xi1eq}, which can be analytically solved:
\begin{equation}
\phi^{(1)}_s(x)=-\frac{1}{6}\frac{\tanh x}{\cosh^2x}\,.
\end{equation}
One can also evaluate the integral \eqref{beta_R} analytically, and the first order
result in $\Or(\epsilon)$ for the reflection coefficient is given as:
\be
\beta_{R}^{(1)}=\frac{2\pi i(4+q^2)}{15\sinh\pi q}\,.\\
\ee
Evaluating the critical amplitude in Eq.\ \eqref{A-crit}
for some values of the perturbation parameter $\epsilon$,
one finds for $\epsilon=0.1$ and $q=1.2$ (corresponding to $\omega\approx2.33$)
$A_{\rm crit}\approx 0.021$ and $A_{\rm crit}\approx 0.041$ for $\epsilon=0.2$.
For increasing values of the frequency $\omega$, $\beta_{R}^{(1)}\to0$ exponentially fast
therefore $A_{\rm crit}$ becomes very small.

In order to confirm the existence of the critical amplitude and compare
its magnitude with the one found in Eq.\ \eqref{A-crit},
we have performed some numerical simulations on the perturbed $\phi^4$ model \eqref{fi6}.
It is not an easy numerical task to measure the critical amplitude since the acceleration
tends to be very small for $A\sim A_{\rm crit}$ and therefore long-time evolution with
large spatial resolution is needed. The value of the perturbation parameter, $\epsilon$
cannot be chosen to be too small either since then $A_{\rm crit}$ becomes so small
that we cannot measure it.
Also the initial conditions used in the simulations
of the perturbed $\phi^4$ theory gave some non-zero contribution to the initial
velocity of the kink.
The measured values of the accelerations were as small as $10^{-9}$, $10^{-8}$.
For $\epsilon=0.1$ and $\omega=2.33$ we have found
that $A_{\rm crit}=0.0168\pm0.0001$, and $A_{\rm crit}=0.0292\pm0.0002$ for
$\epsilon=0.2$.
The measured values for $A_{\rm crit}$ do not agree very precisely with the prediction
of Eq.\ \eqref{A-crit}, the discrepancies being of about $20\%$ resp.\ $30\%$
for $\epsilon=0.1$ resp.\ for $\epsilon=0.2$.
Nevertheless we consider the numerically found values to be consistent
with  Eq.\ \eqref{A-crit} for the following reasons.
$A_{\rm crit}/\epsilon$ still varies considerably (by $\sim13\%$)
for the two considered values of $\epsilon$, indicating that higher order corrections
are still important here.
For the smaller value of $\epsilon$ the discrepancy between the measured value and the result
of Eq.\ \eqref{A-crit} is also smaller.
In any case we have been able to demonstrate the existence of a critical amplitude
in the perturbed $\phi^4$ model above which the radiation pressure becomes negative.
The measured value of $A_{\rm crit}$ is consistent with the theoretical estimate
in Eq.\ \eqref{A-crit}.
Moreover the negative radiation pressure persists for rather large values
of the perturbation parameter, $\epsilon$.
For example in our numerical simulations for $A=0.1$ and $\omega=3.0$
the radiation pressure became positive only for either $\epsilon>1.3$ or $\epsilon<-1.2$.
We have shown that for a rather large range in the magnitude of a generic perturbation of the
$\phi^4$ model, the phenomenon of negative radiation pressure persists.

Finally coming to the perturbation of sG model, we have shown that to order
$\Or(A^4)$ the force is zero. In fact using the analytic
solution of Shin \cite{Shin} we conclude that the force is zero to all orders.
Therefore our derivation for the critical amplitude above which the force acting on
the kink is still negative does not apply to the particular case of the sG model.
We have observed that the kink only oscillates around its initial
position, and the average of its velocity is zero.
This implies, that even a small perturbation of the sG model may change
this qualitative behaviour and determine the motion of the kink.

\section{Conclusions}
We have studied the interaction of a kink in $1+1$ dimensional scalar
models with an incoming wave in perturbation theory.
We have shown that in a certain class of theories
(such as the $\phi^4$ model), the kink is {\sl pulled} towards the
direction of the incident radiation, instead of being pushed back.
This interesting phenomenon constitutes an interesting example
of negative radiation pressure, which in this case is due to the nonlinearities and
to higher order effects.
Comparing the results of the perturbative calculations to numerical simulations
in various field theoretical models (mostly in $\phi^4$ and in sG)
a rather good agreement was found for not too large values of the amplitude of the incoming
wave (up to $A\approx0.2$).
We have also addressed the important problem of structural stability of
the negative radiation pressure with respect to generic perturbations of the theory.
In models where the reflection coefficient is small we have established the existence
of a critical amplitude above which the kink experiences negative radiation pressure.
This is closely related to the robustness of the effect
which has been demonstrated on the example of a perturbation
of the $\phi^4$ model.
We have found in fact that even for
large perturbation there exists a critical amplitude of the incoming wave
above which the radiation pressure becomes negative.

In the sG model the radiation pressure turns out to be zero.
In this model there is an analytical solution corresponding to the superposition
of a kink and an incoming (cnoidal) wave \cite{Shin}, and
we could confirm the correctness of our perturbative results
by comparing them to the expansion of the exact solution.

We have also shown that under the action of the averaged force exerted by the
radiation the kinks accelerate in all these models according to Newton's law.

It is clearly an important open question if the effect of negative radiation pressure is
also present in other, in particular higher dimensional theories.
Our preliminary results suggest that this phenomenon is also present at least
in two other models of quite some physical interest: in
the $2+1$ dimensional complex $\phi^4$ theory (Goldstone's model)
and in the Abelian Higgs model admitting vortices.
This suggests that this effect might not be so rare as one could have
expected at first sight.

The phenomenon of negative radiation pressure is relevant
not only for the interaction of a single kink with radiation but
also for system of many kinks.
One would expect for example that a many-kink system
where the radiation pressure is negative becomes
unstable with respect to collapse.
Since the interaction of well localized kinks is rather weak
(e.g.\ for two $\phi^4$ kinks at a distance $L$ the force between them is
$F\sim\e^{-\alpha L}$), the radiation pressure
can play an important role in many kink systems.

It would clearly be interesting to investigate this effect also for domain walls
(or higher dimensional branes).

\appendix

\section{Details of the calculation of higher order perturbations}\label{appendix:genpert}

In this Appendix we shall present some of the details
of the the 2nd and 3rd order perturbative calculations necessary to find the force acting on the
kink and its acceleration.

All equations arising in perturbation theory are 2nd order linear
inhomogeneous partial differential equations of the form
\begin{equation}\label{eq:gen0}
\ddot\xi^{(n)}+\opL\xi^{(n)}=f^{(n)}(U^{(k)}(\phi_s),\xi^{(l)})
\end{equation}
where the inhomogeneous term must be computed from the solutions
obtained in order lower than $\Or(A^n)$.
To avoid to present too complicated general formulae,
we start by writing out explicitly
the 2nd order equations.
They are obtained by simply
substituting the perturbative expansion for the field $\phi(x,t)$ Eq.\ (\ref{eq:post1})
and for the frequency (\ref{eq:czes}) into the equation of motion (\ref{eq:eq1}):
\begin{equation}
\ddot{\xi}^{(2)}-2\omega^{(0)}\omega^{(1)}\xi^{(1)}+\opL\xi^{(2)}=
-\frac{1}{8}U'''(\phi_s)\biggl(\eta_q^2e^{2i\omega t}-
2\eta_q\eta_{-q}-\eta_{-q}^2e^{-2i\omega t}\biggr)\,. \label{eq:secorder1}
\end{equation}
On the r.h.s. of Eq.\ \eqref{eq:secorder1}
there are two source terms oscillating with frequency $\pm2\omega$
and a time independent term. Therefore we can seek the solutions of Eq.\
\eqref{eq:secorder1} in the form
\begin{equation}
\xi^{(2)}(x,t) = \xi^{(2)}_{+2}(x)e^{2i\omega
t}+\xi^{(2)}_{-2}(x)e^{-2i\omega t}+\xi^{(2)}_{0}(x,t)\,.
\end{equation}
Denoting $m$-th coefficient in Fourier's expansion of $\xi$ of
order $\Or (A^n)$ by $\xi^{(n)}_m$, we obtain the
following equations:
\begin{eqnarray}
-2\omega^{(0)}\omega^{(1)}\xi^{(1)}&=&0,\\
\ddot{\xi}^{(2)}_0 + \Opl\xi^{(2)}_{0}&=&-\frac{1}{4}U'''(\phi_s)\eta_{q}\eta_{-q}\,,
 \label{eq:nzal}\\
\left(\Opl-4{\omega^{(0)}}^2\right)\xi^{(2)}_{\pm2}&=&-\frac{1}{8}U'''(\phi_s)\eta_{\pm q}^2\,.
\label{eq:eq6}
\end{eqnarray}
The first equation gives  immediately $\omega^{(1)}=0$, i.e.\
there is no correction to frequency in the first order.
Projecting Eq.\ \eqref{eq:nzal} onto the translational mode $\eta_t$,
and using the identity
$(\Opl\ddot{\xi}^{(2)}_{0}|\eta_t)=(\Opl\eta_t|\ddot{\xi}^{(2)}_{0})^*=0$.
we obtain
\begin{equation}
(\ddot{\xi}^{(2)}_0|\eta_t)=
-\frac{1}{4}(U'''(\phi_s)\eta_{q}\eta_{-q}|\eta_t)\,. \label{eq:accel1}
\end{equation}
As we shall show now
the r.h.s.\ of equation (\ref{eq:accel1}) vanishes precisely
for reflectionless potentials.
To prove this we shall take the derivative of the eigenvalue problem of the linear operator $\opL$:
\begin{equation}
\left(-\frac{\d^2}{\dx^2}+U''(\phi_s(x))\right)\eta_q(x)=\omega^2\eta_q(x)\,,\label{eq:motion}
\end{equation}
 then multiply it with $\eta_{-q}$ and integrating leads to
\begin{equation}
\int\dx\;\;\eta_{-q}\left(\opL-\omega^2\right)\eta_q'=
-( U'''(\phi_s(x))\eta_{q}\eta_{-q}|\phi_s')\,.
\end{equation}
Integration by parts (over some interval $(-L,L)$) of the l.h.s.
of the above equation and using equation (\ref{eq:motion}) satisfied by $\eta_{-q}$ gives
\begin{equation}
\left.(\eta_q'\eta_{-q}'-\eta_{-q}\eta_q'')\right|_{-L}^{L}=
-( U'''(\phi_s)\eta_{q}\eta_{-q}|\eta_t)\,, \label{eq:boundary}
\end{equation}
where we have also used that the translational mode $\eta_t =\phi_s'$.
Finally it is easy to calculate the boundary values in (\ref{eq:boundary})
by using the asymptotic form for $\eta_q$ (\ref{line1}) and (\ref{line2}) leading
to the interesting identity
\begin{equation}
4q^2|{\mathbf R}|^2=-( U'''(\phi_s)\eta_{q}\eta_{-q}|\eta_t)\,. \label{eq:boundary1}
\end{equation}
This demonstrates that for reflectionless potentials
\ie ${\mathbf R}\equiv0$, $( U'''(\phi_s)\eta_{q}\eta_{-q}|\eta_t)=0$.
Therefore it is fully consistent to assume $\ddot{\xi}^{(2)}_0=0$
in Eq.\ (\ref{eq:nzal}) when ${\mathbf R}\equiv0$.
It is not difficult to obtain the asymptotic form of ${\xi}^{(2)}_0(x)$ since it
has a limit for $x\to\infty$:
\begin{equation}
\xi^{2}_0(x\rightarrow\pm\infty)=-\frac{U'''(\phi_{vac})}{4U''(\phi_{vac})}\,.\label{eq:xi20asy}
\end{equation}
To obtain the asymptotic form
of the solution of equation (\ref{eq:eq6}) we make use of the Green's function
for the inhomogeneous problem. Since all
inhomogeneous equations we have to solve can be written as
\begin{equation}\label{eq:gen1}
\opL\xi^{(n)}_m=f^{(n)}_m(x)\,,
\end{equation}
we give their solution satisfying our boundary conditions in a general form:
\begin{equation}\label{eq:gensol}
\xi^{(n)}_m(x)=-\frac{\eta_{-\kappa}(x)}{W}\int_{-\infty}^x\!\!\dx'\;
\eta_\kappa(x')f^{(n)}_m(x')
-\frac{\eta_{\kappa}(x)}{W}\int^{\infty}_x\!\!\dx'\;\eta_{-\kappa}(x')f^{(n)}_m(x')
\end{equation}
where $\eta_\kappa$ are the solutions of the homogeneous equation (eigenfunctions of $\opL$),
$\kappa=\sqrt{m^2\omega^2-U''(\phi_{vac})}$ is 
the wave number corresponding to the frequency $m\omega$ and
$W=\eta_\kappa\eta'_{-\kappa}-\eta_\kappa'\eta_{-\kappa}=-2i \kappa$
is the Wronskian\footnote{Note that as there is no first order space
  derivative in the equations of motion, the Wronskian is independent
  of $x$, and thus can be calculated from the asymptotic form of
  $\eta_q$ and $\eta_{-q}$.}.
To obtain the asymptotic form of the solutions
we write the integrals as a difference, e.g.:
\begin{equation}
\int_{-\infty}^x\!\!\!\dx'\;\eta_{\kappa} (x')f^{(n)}_m(x')=
\left(\int_{-\infty}^\infty-\int_{x}^\infty\right)\dx'\;
\eta_{\kappa} (x')f^{(n)}_m(x')\,.
\end{equation}
The integral over the real line can be calculated using the method of residua while
the second one can be calculated using the asymptotic form of $\eta_\kappa$  and $f^{(n)}_m$.
This way we obtain the asymptotic form of the solution of Eq.\ (\ref{eq:eq6}):
\begin{equation}\label{eq:xi2asy}
\xi^{(2)}_{+2}(x\rightarrow\pm \infty)=\frac{U'''(\phi_{vac})}{24U''(\phi_{vac})}\eta_{+q}^2+
\alpha_{22,\pm k}(q)\eta_{\mp k}\,,
\end{equation}
where 
$k$ is the wavenumber corresponding to the frequency $2\omega$, and
\begin{equation}
\alpha_{22,k}(q)=-\frac{1}{8W}\int_{-\infty}^\infty\!\!\!\dx'\;\eta_{k}\eta_q^2U'''(\phi_s)\,.
\end{equation}
The solution for the negative frequency can be found as a complex conjugation of the above
solution $\left(\xi^{(2)}_{-2}={\xi^{(2)}_{+2}}^*\right)$.

All the coefficients of the homogeneous part will be denoted as
\begin{equation}
\alpha_{mn,\kappa}(q)=-\frac{1}{W}\int_{-\infty}^\infty\!\!\!\dx'\eta_\kappa(x')f^{(n)}_m(x').
\end{equation}

The third order equations have the following forms:
\begin{gather}
\left(\Opl-9{\omega^{(0)}}^2\right)\xi^{(3)}_{\pm3}=
-\frac{1}{6}U^{\rm (iv)}(\phi_s){\xi^{(1)}_{\pm1}}^3-U'''(\phi_s)\xi^{(1)}_{\pm1}\xi^{(2)}_{\pm2}\,,
\label{eq:eq10a}\\
\left(\Opl-{\omega^{(0)}}^2\right)\xi^{(3)}_{\pm1}=
-U'''(\phi_s)\left(\xi^{(1)}_{\pm1}\xi^{(2)}_{0}+\xi^{(1)}_{\mp1}\xi^{(2)}_{\pm2}\right)-\nonumber\\
-\frac{1}{2}U^{(\rm iv)}(\phi_s){\xi^{(1)}_{\pm1}}^2\xi^{(1)}_{\mp1}+2\omega^{(0)}{\omega^{(2)}}
\xi^{(1)}_{\pm 1}.\label{eq:eq10}
\end{gather}
Taking into account the asymptotic forms of $\xi^{(n)}_m$  equation (\ref{eq:eq10}) can be
rewritten as
\begin{equation}
\begin{split}
\left(\Opl-{\omega^{(0)}}^2\right)\xi^{(3)}_{1}(x\rightarrow\pm \infty)=
-\frac{1}{2}U'''\alpha_{22,\pm k}(q)\eta_{-q}\eta_{\mp k}+\\
+\eta_{+q}\left(\frac{5U'''^2}{48U''}-\frac{U^{\rm (iv)}}{16}+\omega^{(0)}\omega^{(2)} \right),
\end{split}
\end{equation}
where derivatives of $U$ has to be taken at $\phi=\phi_{vac}$.
Note that the l.h.s. of the above equation is the same as for harmonic oscillator with frequency
$q$. On the r.h.s. there is a source term which oscillates with the resonant frequency.
Therefore the following condition must be fulfilled to cancel this resonance term
\begin{equation}
\omega^{(2)}=-\frac{1}{\omega^{(0)}}\left(\frac{5U'''^2}{48U''}-\frac{U^{\rm (iv)}}{16}\right)\,.
\label{eq:omega2}
\end{equation}
This gives the first correction to the frequency. (Note that the values of the derivatives of
the potential must be taken at vacuum.)
Having this we can write the asymptotic form of the eq.\ (\ref{eq:eq10}) in much simpler form
\begin{equation}
\left(\Opl-{\omega^{(0)}}^2\right)\xi^{(3)}_{+1}(x\rightarrow\pm \infty)=
-\frac{1}{2}U'''\alpha_{22,\pm k}(q)\eta_{-q}\eta_{\mp k}
\end{equation}
which leads to the solution
\begin{equation}
\xi^{(3)}_{1}(x\rightarrow\pm \infty)=-\frac{\alpha_{22,k}U'''}{p_{\pm}^2+U''-{\omega^{(0)}}^2}
\eta_{\mp k}\eta_{-q}+\alpha_{31,\pm q}\eta_{\mp q},\label{eq:xi31asy}
\end{equation}
where $p_{\pm}=\mp k-q$.
A computation similar to the above yields
\begin{equation}\label{eq:xi3asy}
\begin{split}
\xi^{(3)}_{3}(x\rightarrow\pm \infty)=\frac{1}{384U''}\left(U''U^{(\rm iv)}+
U'''^2\right)\eta_q^3\mp\\
\mp\frac{1}{2}\frac{U'''\alpha_{22,k}}{p_\pm^2-9{\omega^{(0)}}^2+U''}\eta_{\mp k}\eta_q+
\alpha_{33,\pm s}\eta_{\mp s},
\end{split}
\end{equation}
where 
$s$ is the wavenumber corresponding to the frequency $3\omega$.
As one can see in the third order solution there are only terms which oscillate with time,
therefore its projection onto the translational mode gives no contribution (after averaging
in time) to the time independent part of the acceleration.
The next time independent term appears at 4th order, $\Or(A^4)$.
Then, the equation for $\xi^{(4)}_0$ has the form:
\begin{equation}\label{proj4}
\begin{split}
&\ddot{\xi}^{(4)}_0+\Opl\xi^{(4)}_0=\\
&-\frac{1}{2}U'''(\phi_s)\left(2\xi^{(1)}_{1}\xi^{(3)}_{-1}+
2\xi^{(1)}_{-1}\xi^{(3)}_{1}+2\xi^{(2)}_{2}\xi^{(2)}_{-2}+{\xi^{(2)}_0}^2\right)\\
&-\frac{1}{2}U^{(\rm iv)}
(\phi_s)\left({\xi^{(1)}_1}^2\xi^{(2)}_{-2}+2\xi^{(1)}_1\xi^{(1)}_{-1}\xi^{(2)}_0+
{\xi^{(1)}_{-1}}^2\xi^{(2)}_2\right)\\
&-\frac{1}{4}U^{(\rm v)}(\phi_s){\xi^{(2)}_{1}}^2{\xi^{(2)}_{-1}}^2.
\end{split}
\end{equation}
Now computing the projection of $\xi^{(4)}_0$ to the translational mode, $\eta_t$,
exploiting the obvious reflection symmetries some of the terms in Eq.\ \eqref{proj4}
will not contribute, and we are led to the following result:
\begin{equation}
(\ddot{\xi}^{(4)}_0|\eta_t)=-
\Re(U'''(\phi_s)(\xi^{(3)}_1\eta_{-q}+\xi^{(2)}_2\xi^{(2)}_{-2})+
\frac{1}{4}U^{(\rm iv)}(\phi_s)\xi_2^{(2)}\eta_{-q}^2|\eta_t)\,.\label{eq:accel4}
\end{equation}

\section{Newton's law}\label{appendix:fma}
In this section we show by explicit computation that Newton's law
holds to leading order in PT.
We start by calculating the projection of the second time
derivative of the perturbation $\ddot{\xi}$ onto the translational mode of the kink.
Computing the second order acceleration in PT, by projecting $\ddot\xi$ on the translational mode,
one obtains
\begin{equation}
  \label{eq:ACC2}
  (m a)_2=\int_{-L}^L
  \phi_s' (U'''(\phi_s)\xij{1}{1}\xij{1}{-1})\d x=
  \left.U''(\phi_s)\xij{1}{1}\xij{1}{-1}\right|_{-L}^{L}-\int_{-L}^L
  U''(\phi_s)({\xij{1}{1}}'\xij{1}{-1}+\xij{1}{1}{\xij{1}{-1}}').
\end{equation}
Next looking at the relevant component of the stress-energy tensor,
\begin{equation}
  (-T_{11})_2 = \xij{2}{0}' U'(\phi_s)+U''(\phi_s)\xij{1}{1}\xij{1}{-1}-\xij{1}{1}'\xij{1}{-1}'
  -\dot{\xi}^{(1)}_1\dot{\xi}^{(1)}_{-1}+\phi_s'\xij{2}{0}',
\end{equation}
one sees  that the first and the last terms give negligible contributions  $(-T_{11})|_{-L}^L$, if
$L$ is sufficiently large.
The second term is exactly the boundary term in (\ref{eq:ACC2}). To prove the equality, let us put
\[
-\left.\xij{1}{1}'\xij{1}{-1}'\right|_{-L}^{L}=-\int_{-L}^{L} \xij{1}{1}''\xij{1}{-1}' +
 \xij{1}{1}'\xij{1}{-1}'',
\]
and note that $\dot{\xi}^{(1)}_{1}=i\omega \xi^{(1)}_1$.
Therefore, the remaining terms in $(-T_{11})_2|_{-L}^L$ are equal to
\[
\int_{-L}^L \xij{1}{1}'\left({\ddot{\xi}}^{(1)}_{-1}-\xij{1}{-1}''\right) +
\xij{1}{-1}'\left({\ddot{\xi}}^{(1)}_{1}-\xij{1}{1}''\right)=
               -\int_{-L}^L \xij{1}{1}' U''(\phi_s)\xij{1}{-1} + \xij{1}{-1}' U''(\phi_s)\xij{1}{1},
\]
where we used the equation of motion $\ellop \xij{1}{\pm 1}=0$.
Comparing this with the last term in (\ref{eq:ACC2}) completes the proof.

If the acceleration is vanishing to second order,
the first nontrivial contribution to it can only come form the 4th order.
This is because neither $ma$ nor $T_{11}$ has a zero frequency part in the third order.
The fourth order acceleration is given by $(m_s a)_4 = -\langle
\ddot\xi^{(4)}_0\rangle$ where $\ddot\xi^{(4)}_0$ can be replaced by its source term
(see equation (\ref{proj4})).
This should be equal to $(-T_{11})_4|_{-L}^{L}$, where
\begin{equation}
  \label{eq:T114}
\begin{aligned}
(-T_{11})_4 = & U''(\phi_s)\Big\{ \frac{1}{2}\xij{2}{0}^2 + \xij{2}{2}\xij{2}{-2} +
 \xij{1}{-1}\xij{3}{1}+ \xij{1}{1}\xij{3}{-1}\Big\}
             + U'''(\phi_s)\Big\{ \xij{1}{1}\xij{1}{-1}\xij{2}{0}+\frac{1}{2}\xij{1}{-1}^2\xij{2}{2}+
             \frac{1}{2}\xij{1}{1}^2\xij{2}{-2}\Big\}\\
             -&\frac{1}{2}{\dot\xi}^{(2)}_{0}{}^2-{\dot\xi}^{(2)}_{2}{\dot\xi}^{(2)}_{-2}-
             {\dot\xi}^{(1)}_{-1}{\dot\xi}^{(3)}_{1}-{\dot\xi}^{(1)}_{1}{\dot\xi}^{(3)}_{-1}
             -\frac{1}{2}{\xi'}^{(2)}_{0}{}^2-{\xi'}^{(2)}_{2}{\xi'}^{(2)}_{-2}-
             {\xi'}^{(1)}_{-1}{\xi'}^{(3)}_{1}-{\xi'}^{(1)}_{1}{\xi'}^{(3)}_{-1}\\
             +& U^{(4)}(\phi_s)\frac{1}{4}\xij{1}{1}^2\xij{1}{-1}^2 +
             4\omega\omega^{(2)}\xij{1}{1}\xij{1}{-1}.\\
\end{aligned}
\end{equation}
We remark that to remove the resonance terms from the source of $\xij{3}{\pm 1}$,
one has to introduce a nonzero $q^{(2)}$. This term induces a fourth order correction
to $(T_{11})_2$.
Explicitly it is given by
\[
 (-T_{11})_{\text{4res}}= 4\omega \omega^{(2)}\xij{1}{1}\xij{1}{-1}|_{-L}^L.
\]
To establish Newton's law to this order
proceeds essentially the same way as in the second order case.
Performing the partial integrations in the integral $(ma)_4$ gives all the
terms in the energy-momentum tensor containing the potential as boundary terms.
The integral terms after the partial integration in $(ma)_4$ are
\begin{equation}\label{eq:ACCrem}
\begin{aligned}
-&\int_{-L}^{L}U''(\phi_s)\Big\{
\left(\xij{2}{2}\xij{2}{-2}\right)'+\frac{1}{2}\left(\xij{2}{0}\xij{2}{0}\right)'
+\left(\xij{1}{-1}\xij{3}{1}\right)'+\left(\xij{1}{1}\xij{3}{-1}
\right)' \Big\}\\
-&\int_{-L}^{L}U'''(\phi_s)\Big\{\frac{1}{2}\left(\xij{1}{-1}\xij{1}{-1}\xij{2}{2}\right)'
+\frac{1}{2}\left(\xij{1}{1}\xij{1}{1}\xij{2}{-2}\right)'
+ \left(\xij{1}{1}\xij{1}{-1}\xij{2}{0}\right)'\Big\}\\
-&\int_{-L}^{L}\frac{1}{4} U^{(4)}(\phi_s)\Big( \xij{1}{1}\xij{1}{1}\xij{1}{-1}^2\Big)'.
\end{aligned}
\end{equation}
Note, that the terms in the integrands are the source terms of equations of motion
 (\ref{eq4}), (\ref{eq:nzal}), (\ref{eq:eq6}) and (\ref{eq:eq10}). Replacing them with
the right hand side of these equations of motion, and reorganizing the terms, one gets total derivatives,
the integrals of which are the remaining terms of
$(-T_{11})|_{-L}^L$.

Formula (\ref{eq:T114}) can be used to obtain the momentum in the segment $[-L,L]$ to order $\Or(A^4)$.
The rate of change of the momentum flow is given by Eq.\ \eqref{eq:ZasP} and in fact we
need its integrated form Eq.\ (\ref{eq:noether2}).
Direct substitution of the series (\ref{eq:post1}) and expansion into powers of $A$ shows that
for reflectionless potentials
the first non-vanishing term is of order $\Or(A^4)$, therefore
$\partial_t P=A^4\partial_t P^{(4)}$.
We need to calculate the energy-momentum tensor at the boundaries of the segment
$T_{11}$ proportional to $A^4$, $-T^{(4)}_{11}|_{-L}^{L}$.
It is sufficient to use the asymptotic form of the solutions (\ref{eq:xi20asy}),
(\ref{eq:xi2asy}), (\ref{eq:xi31asy}) and (\ref{eq:xi3asy}) together with the frequency
correction (\ref{eq:omega2}).
The calculation of the boundary term leads to a complicated expression, which
after averaging in time gives the leading term for the force:
\begin{equation}
F=-A^4\left[2k^2\left(|\alpha^2_{22,+k}|-|\alpha^2_{22,-k}|\right)-
4q^2{\rm Re}\,\alpha_{31,-q}\right].\label{eq:ZasP_app}
\end{equation}
At the end of this appendix, we will show
that to leading order the energy of the accelerating kink is
$\frac{1}{2}m v^2$ indeed, as
used in Section \ref{section:force4}.
Writing
$\phi= \phi_s +\xi$, and using Eq.\ (\ref{eq:noether}) we obtain
\be
\epsilon = \frac{1}{2}{\phi_s'}^2 + U(\phi_s) + \phi_s'\xi'+U'(\phi_s)\xi +
\frac{1}{2}{\dot\xi}^2+\frac{1}{2}{\xi'}^2 + \frac{1}{2}U''(\phi_s)\xi^2 +
\frac{1}{6}U'''(\phi_s)\xi^3 + \frac{1}{24}U^{(\rm iv)}(\phi_s)\xi^4\,.
\ee
The terms linear in $\xi$ give no contribution because $\phi_s$ solves the static equation of motion.
The only time-dependent non periodic terms that (up to the fourth order) are those containing
$\xi^{(4)}_0=-\frac{1}{2}a t^2 \phi_s'+\dots$.  In the fourth order, there
are no such terms. In the sixth order, the terms quadratic in $\xi$ give a contribution that is
 proportional to the equation of motion of $\xi^{(2)}_0$ multiplied by
$\xi^{(4)}_0$ (in one term one has to perform a partial integration).
The term $\frac{1}{6}U'''(\phi_s)\xi^3$ gives a term proportional to
 $\xi^{(1)}_1 \xi^{(1)}_{-1}\xi^{(4)}_0\phi_s$, which is antisymmetric, therefore its integral is
0. The last term contains no sixth order contribution proportional to $\xi^{(4)}_0$.
Finally in the eighth order, the quadratic terms give the contribution $\frac{1}{2}m v^2$.

\section{Superposition of the sG kink with a cnoidal wave}\label{appendix:shin}
In this Appendix we review first the remarkable analytic solution corresponding to
the nonlinear superposition of a sG kink and a traveling wave.
This solution has been obtained by H.\ J.\ Shin by Darboux transformation methods \cite{Shin}.
In Ref.\ \cite{Shin} light cone coordinates are used, and the sine-Gordon equation is written as:
\begin{equation}
  \label{eq:sglightcone}
  \partial_{\bar z}\partial_z \tilde\phi=2\beta\sin2\tilde\phi\,,
\end{equation}
where the scale parameter, $\beta$ is kept for bookkeeping purposes as in Ref.\ \cite{Shin}.
In these coordinates the static kink takes the form
\begin{equation}
  \label{eq:kinkLC}
  \tilde\chi=2\arctan e^{-2\sqrt{\beta}(z+\bar{z})}.
\end{equation}
The transformation between the two conventions is:
\begin{equation}
  \label{eq:coord}
  \begin{aligned}
    z=\frac{-x+t}{4\sqrt{\beta}} \quad &\quad\quad\bar{z}=\frac{-x-t}{4\sqrt{\beta}},\\
    x=-2\sqrt{\beta}(z+\bar{z}) \quad &\quad \quad t=2\sqrt{\beta}(z-\bar{z}).
  \end{aligned}
\end{equation}
and
\begin{equation}
  \phi(x,t)=2\tilde\phi(z,\bar{z})\,.
\end{equation}

The solution of interest for our purposes is called type 2 in Ref.\ \cite{Shin},
and it can be written as:
\begin{equation}\label{eq:shinmego}
\partial_z \tilde\phi (z,\bar{z})=2 k \sqrt{\frac{\beta}{V}} \cn(\chi_2,k^2) +
4 \sqrt{\frac{\beta}{V}} {\frac{\cn(u,k^2)}{\sn(u,k^2) ~\dn(u,k^2)}}{\frac{S}{S^2+1}}\,,
\end{equation}
and
\begin{equation}
  \label{eq:sinphi}
  \sin2\tilde\phi=U(1-k^2\sn(\chi_2,k^2)^2)+Vk\sn(\chi_2,k^2)\,,
\end{equation}
where $T=1/(S+1/S)$, $U=4T\left(1-{2T}/{S}\right)$ and $V=-2(1-8T^2)$, and
\begin{equation}\label{eq:chi2}
\chi_2=2\sqrt{\frac{\beta}{V}}(z-V\bar{z})
\end{equation}
and
\begin{equation}
S=  -\frac{{a k \sn(u,k^2) \cn(\chi_2 -u,k^2) \mathcal{X} +
b \dn(u,k^2) \mathcal{Y}}}{{b k \sn(u,k^2) \cn(\chi_2 +u,k^2) \mathcal{Y} +a \dn(u,k^2) \mathcal{X}}},
\label{eq:shinS2}
\end{equation}
with
\begin{equation}
\mathcal{X} = \exp (M \zb+ k N \chi_2) \Theta_t(\chi_2 -u), ~
~\mathcal{Y} = \exp (-M \zb-k N \chi_2) \Theta_t(\chi_2 +u),
\label{eq:3sol2}
\end{equation}
\begin{equation}\begin{aligned}
M &= \sqrt{V \beta} \left[ \frac{\cn (u,k^2)}{ \sn (u,k^2) ~\dn (u,k^2)} +
\frac{\dn (u,k^2) ~\sn (u,k^2)}{\cn (u,k^2)  }
-2 k^2\frac{\sn (u,k^2) ~ \cn (u,k^2)}{ \dn (u,k^2)} \right], \\
N &=\frac{\Theta_t '(u)}{k \Theta_t(u)} +\frac{\cn (u,k^2)}{ 2 k ~\dn (u,k^2) ~\sn (u,k^2) }
-k\frac{\sn (u,k^2) ~\cn (u,k^2)}{\dn (u,k^2)  },
\end{aligned}
\label{eq:MN2}
\end{equation}
where
\begin{equation}
\Theta_t(u) =\theta_4 \left( \frac{\pi u}{2 (K-iK')}\right) = 1 +
2 \sum (-)^n q^{n^2} \cos (\frac{n \pi u}{K-iK'}),
\end{equation}
with $q =\exp [-\pi K' / (K-iK')]$, $K=K(k^2)$ and $K'=K(1-k^2)$.
Here, $\sn$, $\cn$ and $\dn$ denotes Jacobi's elliptic functions, and $K$ is the complete
elliptic integral. We use here the notations and conventions of
Abramowitz and Stegun \cite{AbrSteg}.

We would like to point out that the parameter $k$ in Eq.\ (\ref{eq:shinmego})
should not be confused with the wave number, also denoted by $k$ in the previous
sections of this paper. In fact as it turns out $k=A/2$, and therefore it is
a suitable expansion parameter.
As it stands $\Theta_t(u)$ is not well suited to be expanded for $k\to 0$.
It is more convenient to transform it in a form that is easier to expand for small $k$,
when $q\to0$ for $k\to0$.
In fact applying a modular transformation for the function
$\theta_4(z,q)$, $\Theta_t(u)$ can be brought to the form
\begin{equation}
  \label{eq:modular}
  \Theta_t(u)=e^{i\frac{\pi}{4}}\sqrt{\frac{K-iK'}{K}}\exp\left(\frac{\pi u^2}{4(K-iK')K'}+
  \frac{\pi u^2}{4 K K'}\right)\theta_4\left(\frac{\pi u}{2 K}\left|\tau=\frac{iK'}{K}\right.\right),
\end{equation}
where $q=\exp(i\pi\tau)$ and it is now straightforward to expand the
above function for $k\to0$. Let
\begin{equation}
  \Theta(u)=\theta_4\left(\frac{\pi u}{2 K}\left|\tau=\frac{iK'}{K}\right.\right).
\end{equation}

Since $S$ depends only on $\mathcal{X}/\mathcal{Y}$, the common singular parts can be dropped.
The remaining functions are analytic in $k$. Thus
\begin{equation}
\begin{aligned}
\mathcal{X'} &= \exp (M \zb+ k N' \chi_2) \theta_4\left(\frac{\pi(\chi_2-u)}{2 K}\left|\tau=
\frac{iK'}{K}\right.\right),\\
\mathcal{Y'} &= \exp (-M \zb-k N' \chi_2) \theta_4\left(\frac{\pi(\chi_2+u)}{2 K}\left|\tau=
\frac{iK'}{K}\right.\right),
\end{aligned}
\end{equation}
and
\begin{equation}
  kN'=kN-\left(\frac{-\pi u}{2(K-iK')K'}+\frac{\pi u}{4 KK'}\right)=\frac{\Theta'(u)}{\Theta(u)} +
  \frac{\cn (u,k^2)}{ 2  ~\dn (u,k^2) ~\sn (u,k^2) }
-k^2\frac{\sn (u,k^2) ~\cn (u,k^2) }{ \dn (u,k^2)  }.
\end{equation}

Shin's solution depends on parameters $u$, $k$, $V$ and
$\frac{b}{a}$. As we shall see in the next section,
the perturbative solution depends on the amplitude $A$ of the incident wave, the initial
position $x_0=0$ of the kink and its initial velocity $v_0=0$. In order to obtain the dependence of
Shin's parameters on the parameters of the scattering problem, we'll do a series expansion in $k$.

\subsection{Zeroth order}
The zeroth order part of the functions in $\partial\tilde\phi$ is
\begin{equation}
  M_0=\sqrt{V\beta}\left[\cot u +\tan u\right], \quad\quad\quad (kN')_0=\frac{1}{2}\cot u,
\end{equation}
therefore
\begin{equation}
  \begin{aligned}
    \mathcal{X}'_0 &= \exp\left(\sqrt{V\beta}\tan u \bar{z}+\sqrt{\frac{\beta}{V}\cot u} z\right)=:e^X\\
    \mathcal{Y}'_0 &= \exp\left(-\sqrt{V\beta}\tan u \bar{z}-\sqrt{\frac{\beta}{V}\cot u} z\right)=e^{-X},\\
    S_0            &= -\frac{b\mathcal{Y}'_0}{a\mathcal{X}'_0}=-\frac{b}{a}\exp(-2X),
  \end{aligned}
\end{equation}
thus
\begin{equation}
  \label{eq:shindpzo}
  \partial\tilde\phi_0=-4\frac{\beta}{V}\cot u \frac{\frac{b}{a}e^{-2X}}{1+\frac{b^2}{a^2}e^{-4X}},
\end{equation}
which can be compared to the derivative of a Lorentz boosted kink at $x_0$,
\[\chi_v=4\arctan\exp\left(\frac{x-x_0-vt}{\sqrt{1-v^2}}\right).\]
This comparison shows that
\begin{equation}
  \label{eq:shinparamzo}
  \log\frac{b}{a}=\frac{-x_0}{\sqrt{1-v^2}},\quad\quad\quad\sqrt{V}\tan u=\frac{1+v}{\sqrt{1-v^2}}.
\end{equation}
In what follows, we will look at the case in which the kink is at rest in the origin, thus
$x_0=0$, $v=0$ and therefore $a=b=1$ and $\sqrt V \tan u =1$. Of course, higher order
contributions to these
formulas are possible. Note, that $X=-x/2$.

For the calculation of the zeroth order term in $\sin 2\tilde\phi$, one needs
\begin{equation}
  \label{eq:U0}
  U_0=\frac{4 S_0(S_0^2-1)}{(1+S_0^2)^2},
\end{equation}
therefore
\begin{equation}
  \label{eq:sin2phi0}
  (\sin2\tilde\phi)_0=U_0=-\frac{2\tanh x}{\cosh x},
\end{equation}
in which we have already plugged in the above parameters. This is in agreement with
$\tilde\phi_0=\tilde\chi=2\arctan e^{x}$.

\subsection{First order waveform}
$M$ and $kN'$ do not have first order terms. Therefore, neither do $\mathcal{X}$ and
$\mathcal{Y}$. Thus
\begin{equation}
  \label{eq:shindpfo}
  \partial_z\tilde\phi=2k\sqrt{\frac{\beta}{V}}-2\sqrt{\frac{\beta}{V}}\frac{1}{\cosh2X}-
  2k\cot u\frac{\tanh2X}{\cosh2X}\left[e^X\sin u \cos(\chi_2-u)-e^{-2X}\sin u\cos(\chi_2+u)
\right],
\end{equation}
which can be written as
\begin{equation}
  \label{eq:shindpfoeta}
  \partial\tilde\phi=-\sqrt\beta\frac{2}{\cosh2X}+\frac{k}{2}\left[\partial\eta_qe^{i\omega t}+
  \partial\eta_{-q}e^{-i\omega t}\right],
\end{equation}
provided that $\chi_2=\omega t+qx$, i.e.
\begin{equation}
  \label{eq:shinparamfo}
  \sqrt{V}=\omega+q\quad\quad\quad \frac{1}{\sqrt{V}}=\omega-q.
\end{equation}
Thus
\begin{equation}
  \phi=\chi+2k\frac{1}{2}\left(\eta_qe^{i\omega t}+\eta_{-q}e^{-i \omega t}\right).
\end{equation}
Here we can see, that the amplitude is
\begin{equation}
  \label{eq:amplitude}
  A=2k.
\end{equation}
Similarly, $\sin2\tilde\phi$'s first order term is
\begin{equation}
  \label{eq:sinsin2phifo}
  (\sin2\tilde\phi)_1=U_1+V_0\sin\chi,\quad U_1=-\frac{4(1-6S_0^2+S_0^4)S_1}{(1+S_0^2)^3},\quad V_0=
  -2(1-8T_0^2),
\end{equation}
and then, with the right choice of the signs of the square roots,
\begin{equation}
  \label{eq:shinpfo}
  \tilde\phi_1=\frac{(\sin2\tilde\phi)_1}{2\sqrt{1-(\sin2\tilde\phi)_0^2}}=\frac{\cosh^2x}{\cosh2x-3}(\sin2\tilde\phi)_1,
\end{equation}
which can be simplified to
\begin{equation}
  \label{eq:shinpfosimp}
  \phi_1=\eta_qe^{i\omega t}+\eta_{-q}e^{-i\omega t}.
\end{equation}

\subsection{Second order waveform}
Here, one has to expand $T=T_0+k T_1+k^2T_2$. The resulting formulae are
\begin{equation}
  \label{eq:texpand}
  T_0=\frac{1}{\frac{1}{S_0}+S_0},\quad T_1=\frac{1-S_0^2}{(1+S_0^2)^2}S_1,\quad T_2=
  \frac{S_0(S_0^2-3)S_1^2+(1-S_0^4)S_2}{(1+S_0^2)^3}.
\end{equation}
Let $4\sqrt{\beta/V}f(u)=4\sqrt{\beta/V}\cn(\chi_2,k^2)/(\sn(u,k^2)\dn(u,k^2))$ denote
the coefficient of $T$ in $\partial\tilde\phi$. Then
\begin{equation}
  \label{eq:shindpso}
  \partial\tilde\phi_2=4\sqrt\frac{\beta}{V}(f_2 T_0+f_0 T_2),
\end{equation}
where
\[
\begin{aligned}
  f_0 &= \cot u,\\
  f_2 &= \frac{4u-\sin4u}{16\sin^2 u}.
\end{aligned}
\]
Similarly,
\begin{equation}
  \label{eq:shinsin2pso}
  (\sin2\tilde\phi)_2=U_2-2U_0\sin^2\chi_2 +V_1\sin\tilde\chi,
\end{equation}
where
\begin{equation}
  U_2=\frac{4S_0(9-14S_0^2+S_0^4)S_1^2-4(1-5S_0^2-5S_0^4+S_0^6)S_2}{(1+S_0^2)^4},\quad\quad
  V_1=32T_0T_1.
\end{equation}
Again, with the right choice of the signs in the square roots,
\begin{equation}
  \label{eq:shinpso}
  \tilde\phi_2=-\frac{(\sin2\tilde\phi)_0(\sin2\tilde\phi)_1^2}{4(1-(\sin2\tilde\phi)_0^2)^{3/2}}-
  \frac{(\sin2\tilde\phi)_2}{2(1-(\sin2\tilde\phi)_0^2)^{1/2}},
\end{equation}
which can be simplified to get
\begin{equation}
  \label{eq:shinpsosimp}
  \begin{aligned}
\tilde\phi_2 = &-t\frac{4u+\sin4u}{8\sin 2 u}\frac{1}{\cosh x}+e^x\left(\frac{2}{\cosh^2 x}-1\right)\\
         &\times\frac{\left(2e^x x \cosh x +\cos(\omega t-qx)\sin(2u)
             \left(e^{2x}\sin(\omega t-qx-2u)+\sin(\omega t-q x +2u)\right)\right)}
             {1-6e^{2x}+e^{4x}}.
  \end{aligned}
\end{equation}
the first term shows that the kink moves with a constant velocity of
\[v=-k^2\frac{4u-\sin4u}{8\sin 2 u},\]
which can be cancelled by adding a correction to $\sqrt{V}\tan u$, i.e.
\begin{equation}
  \sqrt{V}\tan u = 1-k^2\frac{4u-\sin4u}{8\sin 2 u}.
\end{equation}
The change in the first zeroth order term caused by the $k^2$ part of the parameters is of
the second order in $k$. It cancels the above constant velocity.

Because of $A=2k$, $A^2=4 k^2$, the second order part of the solution in the usual form is the
half of the above $\phi_2$, which, by introducing $q$ instead of $u$ and $\omega$
can be further simplified:
\begin{equation}
  \label{eq:so2}
\begin{aligned}
  \xi^{(2)}=\xij{2}{2}e^{2i\omega t}+\xij{2}{-2}e^{-2i\omega t}+\xij{2}{0}=
  \frac{e^{i(2\omega t+2qx)}(iq-\tanh x)}{16\cosh x(1+q^2)}&+
  \frac{e^{-i(2\omega t+2qx)}(-iq-\tanh x)}{16\cosh x(1+q^2)}\\
  &+\frac{1}{8\cosh x}\left(\frac{\tanh x}{1+q^2}-2x\right)\,.
\end{aligned}
\end{equation}
which is exactly the same as (\ref{eq:so}), the solution obtained using the perturbative method.

\subsection{Motion of the kink}
The time-averaged (i.e.\ zero frequency part) motion of the zero of the kink can be also
calculated in an exact form. The condition is $\phi=0$, thus $\sin\phi=0.$ This
can be written as
\begin{equation}
M\bar{z}+kN'\chi_2=0.
\end{equation}
This leads to
\begin{equation}\label{eq:sGkinkv}
\frac{v+1}{v-1}=\frac{z}{\bar{z}}=-\frac{M-kN'2\sqrt{V\beta}}{2\sqrt{\frac{\beta}{V}}kN'}.
\end{equation}
This means that the velocity of the kink is a constant $v$, which is a free parameter of
Shin's solution.

\section*{ACKNOWLEDGMENTS}
We thank Professor H.~J.~Shin for making the Mathematica notebook file with
the computations for his paper \cite{Shin} available to us, and
Professor Z.~Horv\'ath for useful discussions and advices. This
research has been supported by the OTKA grant 61636, and the ESF
Program ``COSLAB''.

\end{document}